\documentclass[useAMS,usenatbib]{mn2e}

\usepackage{graphicx}
\usepackage{natbib}
\usepackage{longtable}

\usepackage{txfonts}
\title[Herschel Virgo Cluster]{The Herschel Virgo Cluster Survey XVI: a cluster inventory\thanks{{\it Herschel} is an ESA space observatory with science instruments provided by European-led Principal Investigator consortia and with important participation from NASA.}}
\author[Davies et al.]
{J. I. Davies$^{1}$,
S. Bianchi$^{2}$,
M. Baes$^{3}$, 
G. J. Bendo$^{4}$,
M. Clemens$^{5}$,
I. De Looze$^{3}$, \newauthor
S. di Serego Alighieri$^{2}$,
J. Fritz$^{3}$, 
C. Fuller$^{1}$,
C. Pappalardo$^{6}$, 
T. M. Hughes$^{3}$,
\newauthor
S. Madden$^{7}$,
M. W. L. Smith$^{1}$, 
J. Verstappen$^{3}$ and
C. Vlahakis$^{8}$. \\ 
$^{1}$School of Physics and Astronomy, Cardiff University, The Parade, Cardiff, CF24
3AA, UK. \\
$^{2}$INAF-Osservatorio Astrofisico di Arcetri, Largo Enrico Fermi 5, 50125 Firenze, Italy. \\
$^{3}$Sterrenkundig Observatorium, Universiteit Gent, Krijgslaan 281 S9, B-9000 Gent,
Belgium. \\
$^{4}$Jodrell Bank Centre for Astrophysics, School of Physics and Astronomy, 
University of Manchester, Oxford Road, Manchester M13 9PL, UK. \\
$^{5}$INAF-Osservatorio Astronomico di Padova, Vicolo dell'Osservatorio 5, 35122 Padova,
Italy. \\
$^{6}$
CAAUL, Observat\'orio Astron\'omico de Lisboa, Universidade de Lisboa, 
Tapada da Ajuda, 1349-018, Lisboa, Portugal. \\
$^{7}$Laboratoire AIM, CEA/DSM- CNRS - Universit\'e Paris Diderot, Irfu/Service, Paris, France. \\
$^{8}$Joint ALMA Observatory / European Southern Observatory, Alonso de Cordova
3107, Vitacura, Santiago, Chile.
 }

\begin{document}

\date{Original January 2011}


\maketitle


\begin{abstract}
Herschel FIR observations are used to construct Virgo cluster galaxy luminosity functions and to show that the cluster lacks the very bright and the numerous faint sources detected in field galaxy surveys. The far-infrared SEDs are fitted to obtain dust masses and temperatures and the dust mass function. The cluster is over dense in dust by about a factor of 100 compared to the field. The same emissivity ($\beta$) temperature relation applies for different galaxies as that found for different regions of M31. We use optical and HI data to show that Virgo is over dense in stars and atomic gas by about a factor of 100 and 20 respectively. Metallicity values are used to measure the mass of metals in the gas phase. The mean metallicity is $\sim0.7$ solar and $\sim$50\% of the metals are in the dust. For the cluster as a whole the mass density of stars in galaxies is 8 times that of the gas and the gas mass density is 130 times that of the metals. We use our data to consider the chemical evolution of the individual galaxies, inferring that the measured variations in effective yield are due to galaxies having different ages, being affected to varying degrees by gas loss. Four galaxy scaling relations are considered: mass-metallicity, mass-velocity, mass-star formation rate and mass-radius - we suggest that initial galaxy mass is the prime driver of a galaxy's ultimate destiny. Finally, we use X-ray observations and galaxy dynamics to assess the dark and baryonic matter content compared to the cosmological model.
\end{abstract}


\section{Introduction} 

\begin{figure*}
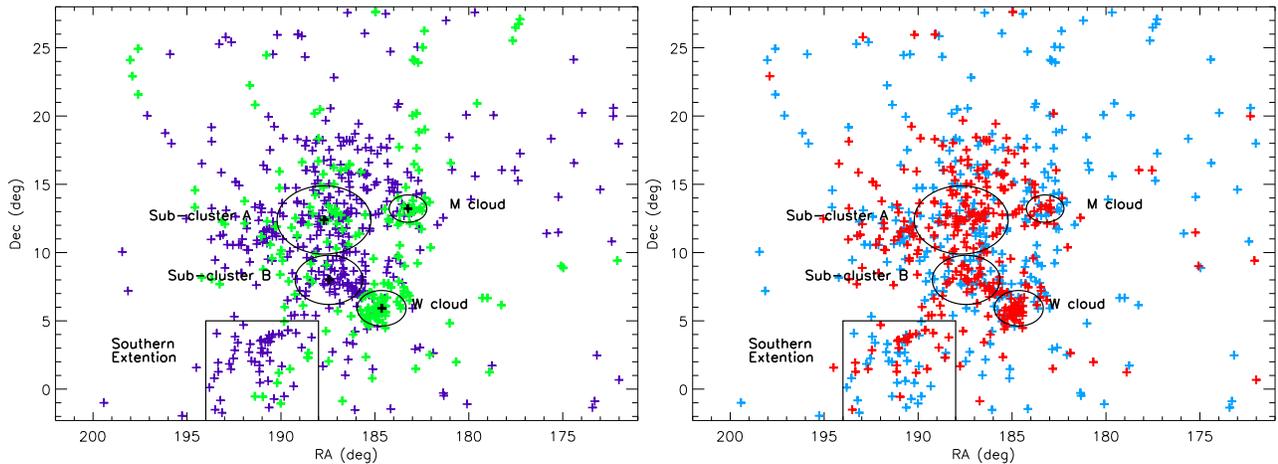

\centering
\includegraphics[scale=0.52]{cluster_clouds.epsi}
\includegraphics[scale=0.52]{cluster_col.epsi}
\caption{Data from a $30\times30$ sq deg region centred on the Virgo cluster and extracted from the SDSS spectroscopic data. Left - galaxies with $400 <v_{Helio}<1600$ km s$^{-1}$ are shown in purple and those with $1600 <v_{Helio}<2659$ km s$^{-1}$ are shown in green. Right - galaxies with $(g-r)>0.62$ are shown in red and those with $(g-r)<0.62$ in blue. The circles and box indicate the various substructures discussed in the text.} 
\end{figure*}

\begin{figure*}
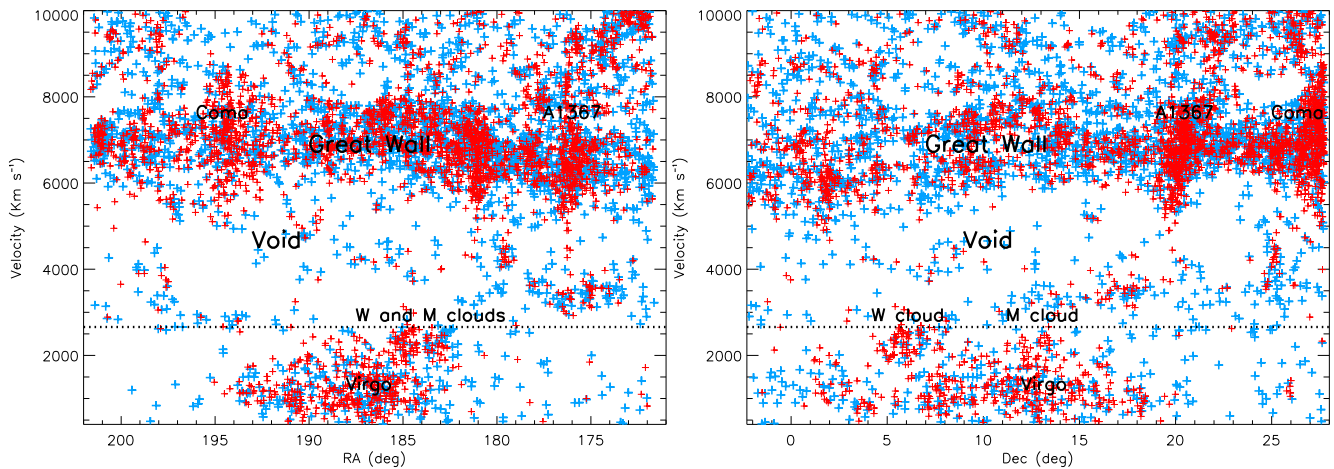

\centering
\includegraphics[scale=0.52]{sdss_vel_ra.epsi}
\includegraphics[scale=0.52]{sdss_vel_dec.epsi}
\caption{Left - Right ascension against velocity for all SDSS spectroscopic data galaxies within 10,000 km s$^{-1}$ over a $30\times30$ sq deg region centred on the Virgo cluster. Galaxies with $(g-r)>0.60$ are shown in red and those with $(g-r)<0.60$ in blue. Right - as figure on the left except this is now declination against velocity.  The black dotted line indicates our velocity limit ($v_{Helio}<2659$ km s$^{-1}$) for the Virgo sample. } 
\end{figure*}

At 17-23 Mpc the Virgo cluster is the nearest large grouping of galaxies to us. It has played a prominent role in astronomical research since the identification of an excess of nebulous objects in this area of sky was noted by both Messier and Herschel. It was first recognised as a group of extra-galactic objects by Shapley and Ames (1926).  The proximity of the Virgo cluster enables us to study both the general properties of galaxies and the way in which the cluster environment may have affected how galaxies evolve. The cluster contains a wide morphological mix of galaxies that subtend some of the largest angular sizes for extra-galactic objects (The elliptical M87 subtends 7 arc min, and the spiral M58 subtends 6 arc min diameter) and so allows us to study not only large numbers of galaxies, but also individual galaxies in detail.

Recent surveys of the Virgo cluster include X-ray (Bohringer et al. 1994), Ultra-violet (Boselli et al. 2011), optical (VCC, Binggeli et al. 1985, ACS, Cote et al. 2004, VGVS, Mei et al. 2010, SDSS, Abazajian et al. 2009), near infrared (2MASS, Skrutskie et al. 2006, UKIDSS, Warren et al. 2007), far-infrared (IRAS, Neugebauer et al. 1984, Herschel, Davies et al. 2012) and 21cm, (ALFALFA, Giovanelli et al. 2005, VIVA, Chung et al. 2009, AGES, Taylor 2010). Prominent amongst these surveys is the optical survey of Binggeli et al. (1985), which listed about 2000 cluster members and has subsequently served as the primary input for many of the other surveys. Clearly, no other galaxy cluster has its galaxy population known to this level of detail and for this reason we now intend to put these various data sources together to hopefully obtain a clearer picture of what constitutes the Virgo cluster.

Before starting on this inventory we will set the scene by reviewing the cluster's structure and its environment. None of this discussion will be very new, but much of what was done in the past was drawn from various, numerous and very different data sources (Binggeli et al. 1987, Binggeli et al. 1993, Gavazzi et al. 1999) whereas we can now use the uniform Sloan Digital Sky Survey (SDSS) spectroscopic data (Strauss et al. 2002) to review the conclusions drawn by others. 

The SDSS spectroscopic sample consists of spectra of all SDSS objects with a $g$ band magnitude of $\le 17.7$. From this extensive data set we have selected all those objects that have been assigned a velocity of between 400 and 10,000 km s$^{-1}$ within a $30 \times 30$ sq deg region centred on M87 (RA(J2000)=187.706, Dec(J2000)=12.391). The minimum value of 400 km s$^{-1}$ was chosen to avoid confusion with Galactic stars and the maximum value of 10,000 km s$^{-1}$ is arbitrary, but we wanted to show the larger scale structure around Virgo. 
\footnote{Note: - velocities below 400 km s$^{-1}$ are only exclude here so that the structure of Virgo can be seen in the figures and is not confused by other non-Virgo SDSS objects. There are of course Virgo galaxies with velocities less than 400 km s$^{-1}$, which will be included in all of the quantitative analysis described later.}
We will adopt the distances to the various components of the cluster as given in Gavazzi et al. (1999). They use Tully-Fisher and fundamental plane scaling relations to obtain velocity independent distances to many of the brighter cluster galaxies. 

In Fig. 1 we show the distribution of cluster galaxies on the sky by limiting the velocity range to a maximum of $\sim2700$ km s$^{-1}$. As we will show below this is the highest velocity of any of the galaxies in our Herschel sample and clearly, as shown in Fig. 2, this separates the cluster in velocity, from other galaxies in other galactic structures. In Fig. 1 (left) we show the position of the galaxies separated into two velocity ranges, $400 <v_{Helio}<1600$ km s$^{-1}$ (463 galaxies) are shown in purple and those with $1600 <v_{Helio}<2659$ km s$^{-1}$ (227 galaxies) are shown in green. Fig. 23, discussed in more detail below, shows that the Virgo cluster galaxy velocity distribution is double peaked and that it roughly corresponds with the above two intervals. Following Binggeli et al. (1987) we identify five structures on Fig. 1 (left):
\begin{enumerate}
\item Sub-cluster A - Galaxies around, but not exactly centred on M87, which has a velocity of about 1300 km s$^{-1}$. Binggeli et al. (1987) describe sub-cluster A as rich in early type galaxies. Gavazzi et al. (1999) place sub-cluster A at 17 Mpc, a distance we will use in what follows.
\item Sub-cluster B - Galaxies around, but not exactly centred on M49, which has a velocity of about 1000 km s$^{-1}$. Binggeli et al. (1987) describe sub-cluster B as rich in late type galaxies and say that it is falling into sub-cluster A from behind. Gavazzi et al. (1999) place sub-cluster B at 23 Mpc, again a distance we will use in what follows.
\item W cloud - Galaxies in this region seem to be isolated spatially and have a greater velocity ($\sim$2200 km s$^{-1}$) than the sub-clusters. Binggeli et al. (1987) say that the distance to the W cloud is about twice that to the sub-clusters and that the W cloud is falling into the sub-clusters.
\item M cloud - Galaxies in this region again seem to be isolated spatially and have a greater velocity ($\sim$2200 km s$^{-1}$) than the sub-clusters. Binggeli et al. (1987) say that the distance to the M cloud is again about twice that to the sub-clusters and that the M cloud is falling into the sub-clusters.
\item Southern extension - there is a filamentary structure that extends to the south of the cluster. Galaxies in the southern extension are at about the same distance as the sub-clusters and Binggeli et al. (1987) say that they also are falling into the sub-clusters.
\end{enumerate}
Fig. 1 (right) is the same as on the left except that now the colour coding picks out intrinsically red and blue galaxies. We simply divided the sample of 690 galaxies in half at the median (g-r) colour of 0.62. Crudely associating red with early and blue with late types we can see the morphology density relation (Dressler, 1980) with red galaxies more concentrated into the identified structures - the exception being the southern extension. Both of the clouds seem to contain their fair share of red galaxies indicating that whatever processes that give rise to the morphology density relation operate on the scale of the clouds as well as on the scale of the sub-clusters.

To reveal the environment of the Virgo cluster, in Fig. 2 we show the full data set out to 10,000 km s$^{-1}$ in position velocity plots. Virgo (sub-clusters A and B) along with the W and M clouds are clearly distinguished in velocity because of the approximate 3,000 km s$^{-1}$ void that lies behind Virgo. The black dotted line is at $v \approx 2700$ km s$^{-1}$ and clearly shows why we used this velocity to isolate Virgo cluster galaxies in Fig.1. At about 6,000 km s$^{-1}$ we find what has become known as the Great Wall, which is a huge filament of galaxies stretching across the sky. Contained within the Great Wall are two other well known clusters Coma and A1367. In Fig. 2 we have again distinguished galaxies via their colour, splitting the sample of 5355 galaxies in half at (g-r)=0.60. Again the morphology density relation is apparent on these much larger scales 

Having hopefully set the scene by briefly describing the structure of the cluster and the environment it resides within, we will now go on and consider the properties of the cluster and its constituent galaxies in much more detail. We will initially concentrate on the far-infrared properties of the galaxies as measured by the Herschel Observatory. We will then add to this new data from the SDSS, so that we can consider the stellar properties of the galaxies, and from the Arecibo Legacy Fast ALFA (ALFALFA) survey to consider the gas properties. In addition and where required we will add information on X-ray gas, warm/hot gas, star formation rates and molecular gas to try and get as complete picture as possible of the cluster and its galactic population. We will then use this data to consider the chemical evolution of the galaxies and the cluster, its current star formation rate, galaxy scaling relations (mass, size, radius) and the total mass including the X-ray gas and dark matter. Finally, we will briefly compare the properties of the two sub-clusters A and B described above.

This paper is a continuation of a series of papers written by us, primarily using data taken from our Herschel guaranteed open time project the Herschel Virgo Cluster Survey (HeViCS). In the previous 13 papers we have described: the survey  and considered the properties of the bright galaxies in a single central 4$\times$4 sq deg field (paper I, Davies et al., 2010), how the cluster environment truncates the dust discs of spiral galaxies (paper II, Cortese et al., 2010), the dust life-time in early-type galaxies (paper III, Clemens et al., 2010), the spiral galaxy dust surface density and temperature distribution (paper IV, Smith et al., 2010), the properties of metal-poor star-forming dwarf galaxies (paper V, Grossi et al., 2010), the lack of thermal emission from the elliptical galaxy M87 (paper VI, Baes et al., 2010), the far-infrared detection of dwarf elliptical galaxies (paper VII, De Looze et al., 2010), the properties of the 78 far-infrared brightest cluster galaxies (paper VIII, Davies et al., 2012), the dust-to-gas ratios and metallicity gradients in spiral galaxies (paper IX, Magrini et al., 2011), the cold dust molecular gas relationship (paper X, Corbelli et al., 2012), the environmental effects on molecular gas and dust (paper XI, Papalardo et al., 2012), the far-infrared properties of 251 optically selected galaxies, (paper XII, Auld et al., 2013) and the dust properties of early type galaxies (paper XIII, di Serego Alighieri et al., 2013). A further five papers (Boselli et al. 2010, Cortese et al. 2012, Boselli et al. 2012, Smith et al. 2012, 2012a) discuss HeViCS galaxies together with  other galaxies observed as part of the Herschel Reference Survey (HRS). 

\section{Data}
We use as our starting point the Herschel data presented and described in Auld et al. (2013). This consists of observations of a total area of 84 sq deg made using Herschel in parallel scan map mode to obtain data in five bands (100, 160, 250, 350 and 500$\mu$m). For a full discussion of this data and its reduction and calibration we refer the reader to Davies et al. (2012) and Auld et al. (2013). 

Using the fully reduced Herschel data we then used the optically selected Virgo Cluster Catalogue (VCC, Binggeli et al. 1985) as the basis of a search for far-infrared emission from VCC members present in the Herschel data. The resulting galaxy sample is fully described in Auld et al. (2013), including a list of flux densities, dust masses and temperatures. Here, we give a brief summary. We used an automated routine to search for far-infrared emission at the position of the 750 VCC galaxies in our Herschel survey area. This resulted in the detection at 250$\mu$m of 251 galaxies. Although it is not ideal to have an optical rather than a far-infrared selected sample we have no other way of ensuring that we have a pure cluster sample rather than one contaminated by background sources. We will show below that there is no evidence for additional cluster far-infrared sources missed by our selection method. 

The 251 galaxies listed by Auld et al., (2013) extend in distance (as given in the GOLDMINE database, Gavazzi et al. 2003) from 17 to 32 Mpc with galaxy groupings at 17, 23 and 32 Mpc. This range of 15 Mpc in depth is large for a cluster and much larger than the linear size we survey on the plane of the sky (about 4 Mpc at a distance of 23 Mpc). For this reason in this paper we restrict our analysis to galaxies with distances of 17 and 23 Mpc so that line-of-sight and plane of sky distances are comparable. These distances correspond with those of sub-cluster A containing M87 and sub-cluster B containing M49 (Gavazzi et al. 1999) - as described in the introduction. It excludes galaxies in the clouds and the southern extention. This leads to a surveyed volume of about 62.4 Mpc$^{3}$. \footnote{The area of sky covered by the Herschel observations described here is about a factor of 1.3 larger than that used in Davies et al. (2012) because the area of sky covered by the full eight scan data set is larger.}
This distance scale is consistent with that recently measured by Mei et al. (2007) and with a Hubble constant of 73 Km s$^{-1}$ Mpc$^{-1}$, which we will use where required throughout this paper.

Restricting distances to between 17 and 23 Mpc leads to a sample of 208 galaxies. However, upon inspection of the data in GOLDMINE 4 of these were discovered to be listed as background galaxies (VCC12, VCC28, VCC40 and VCC262). A further galaxy VCC881(North) was removed because of its close proximity to VCC881(south) and hence possible confusion, this gives 203 galaxies. As described above SDSS velocity data is now available for a flux limited ($g \le 17.7$) sample of galaxies over the Virgo cluster region. To see if we were missing any optical sources not included in the VCC (which was selected to almost equivalently $B \approx 18.0$) we searched the SDSS archive for galaxies over our survey area, which had a helio-centric velocity of 400-2600 km s$^{-1}$ - this led to 43 new optical detections. Four of these new optical detections were subsequently found to have been detected in our Herschel data at 250$\mu$m and these have been added to our list to make a total of 207 galaxies in our final sample - 147 at 17 and 60 at 23 Mpc. This gives a mean distance for our sample of 18.7 Mpc.\footnote{We assume the new detections are at this mean distance of 18.7 Mpc.}

In order to carry out our inventory of the cluster we require, in addition to our Herschel data, information at other wavelengths. We have used (where available) B and H band magnitudes from the NASA Extra-galactic Database (NED), 21cm atomic hydrogen observations from the Arecibo Legacy Fast ALFA Survey (ALFALFA, Giovannelli et al. 2007), used molecular hydrogen masses from Corbelli et al. (2012) and Young et al. (2011) and data from the compilation of galaxy metalicities and star formation rates from Hughes et al. (2013). It is our intention to use correlations within these sparsely populated data sets to obtain the data we require for all 207 of our Herschel sample objects.

\section{Luminosity functions}
The Auld et al. (2013) data is obtained from the positions of optically selected galaxies in the VCC. This is not ideal as we would prefer to select galaxies via their far-infrared flux density when constructing far-infrared luminosity functions. The problem of course is identifying which far-infrared sources belong to the cluster and which are in the background. To address this issue we have carried out a faint galaxy number count analysis of the Virgo field and compared it to the faint galaxy number counts derived from the North Galactic Pole (NGP) field observed by the H-ATLAS consortium (Eales et al. 2010). We have done this at 250$\mu$m because this is the wavelength at which the Auld et al. (2013) sample was selected i.e. all the objects had to have a 250$\mu$m detection. Our primary motivation for doing this is to compare the number counts from the general field (NGP) with that obtained by 'looking through' the Virgo cluster into the Universe beyond. In this way we can assess the contribution made by the cluster to the total galaxy counts.

The NGP data is fully described in Valiante et al. (in prep). To summarise it consists of two orthogonal scans covering $\sim$180 sq deg of sky compared to the eight scans used for the Virgo data over $\sim$85 sq deg. The NGP data has been reduced using the same data reduction pipeline as the Virgo data, with the exception that the NGP data has been gridded onto a 5 arc sec pixel scale while the Virgo data uses 6 arc sec pixels. We have used the source detection programme Sextractor to extract faint sources from both the Virgo and NGP data, taking great care to apply identical methods to both fields. This is important because it quickly became apparent that very small changes in, for example the threshold detection level, can greatly alter the number of sources detected.

The Virgo field suffers from considerable cirrus contamination, which we substantially reduced by subtracting from the data a smoothed version of itself. The cirrus occurs on many spatial scales, but by subjectively considering the obvious features we decided on Gaussian smoothing with a full width half maximum of 11 arc min to create the smoothed frame. As the Herschel data has a mean value of zero (by design) the subtraction conserves flux in the image. The subtraction 'flattens' the sky considerably, but does lead to some dark rings (negative values) around the brighter galaxy images because of their influence when deriving the local sky value. This should not be a problem as bright sources are least affected by sky brightness errors and we are primarily interested in the possibility of there being faint (relatively small) galaxies in Virgo that we have missed via our optical source selection. We will show below that at bright magnitudes the number counts are consistent with the Auld et al. (2013) data. The above does not rule out the possibility of there being rather large ($\sim$10 arc min or above) sized far-infrared sources in Virgo - these would be impossible to distinguish from the cirrus using our method.

\begin{figure}
\centering
\includegraphics[scale=0.52]{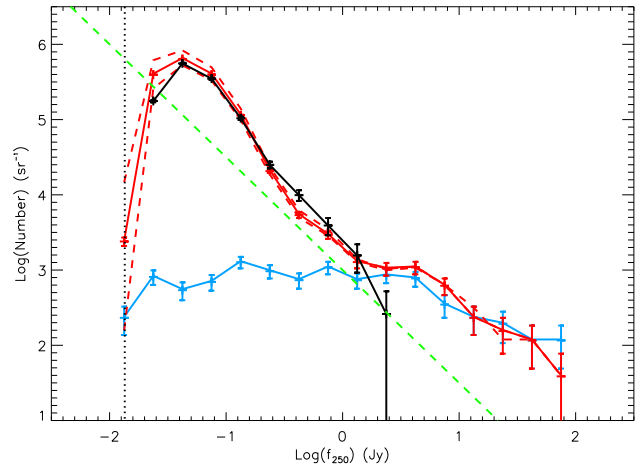}
\caption{The Virgo field 250$\mu$m faint galaxy number counts (solid red line) compared to the faint galaxy number counts in the general field as measured in the H-ATLAS NGP data (solid black line). The red dashed lines indicate lower and upper level expectations for the Virgo counts given the variations in sky noise in the Virgo field, which suffers from larger Galactic cirrus contamination. The blue line shows the number counts using the 251 galaxies from the Auld et al. (2013) data. The green dashed line has a slope of 1.5 and indicates the expected counts for a non-evolving Euclidean universe. The black dotted line indicates the minimum flux density detectable given the minimum detection area and 1$\sigma$ noise level in the NGP data.}
\end{figure}

Although the NGP field seems to be mostly devoid of Galactic cirrus, to be consistent we applied the same sky subtraction method to the NGP data. For a realistic comparison to be made between the two fields we need to correctly account for the different noise characteristics of the two fields and see how this affects what Sextractor does. Sextractor measures the background noise over a specified area (mesh size) and uses this 'local' value to extract sources of a given size above some multiple of this noise. We chose the mesh size to be 100 sq arc min, much larger than a typical galaxy. We then had to determine the fraction ($f$) of the background noise to set the detection threshold at, given the noise in the two data sets (NGP and Virgo). We experimented with multiple 100 sq arc min apertures to see how different the noise was both within and between the two data sets. The noise level was lower in the longer exposure Virgo data (but by only a factor of 0.9), but there was also a larger scatter in the values, presumably due to the residual Galactic cirrus (see below). We set the detection threshold at 1$\sigma$ in the NGP field, which equated to a 1.6$\sigma$ detection in the Virgo data with its larger pixels and lower noise per pixel. We set the minimum detection size to that of the 250$\mu$m beam area of 423 sq arc sec.

The derived number counts for both the NGP and Virgo fields are shown in Fig. 3. The solid black line shows the NGP number counts. The counts are consistent with other counts and detection methods used on previous H-ATLAS data (Clements et al. 2010). The green dashed line has a slope of 1.5, which is what is expected for a non-evolving  Euclidean universe. The line has been drawn to illustrate the sharp rise in the counts above a 'flat non-evolving' model at a flux density of about 0.2 Jy, as previously shown by Clements et al. (2010). The black dotted line indicates the minimum possible flux density for a source the size of the beam with all pixels at 1$\sigma$ above the background (0.013 Jy). The solid red line shows the counts in the Virgo field. The excess of bright cluster sources is clearly seen departing from the background at about 1 Jy. The blue line is for the original 251 galaxies from the Auld et al. (2013) sample and is consistent with the bright galaxy data ($> 1$ Jy) obtained using Sextractor (solid red line). 

\begin{figure}
\centering
\includegraphics[scale=0.52]{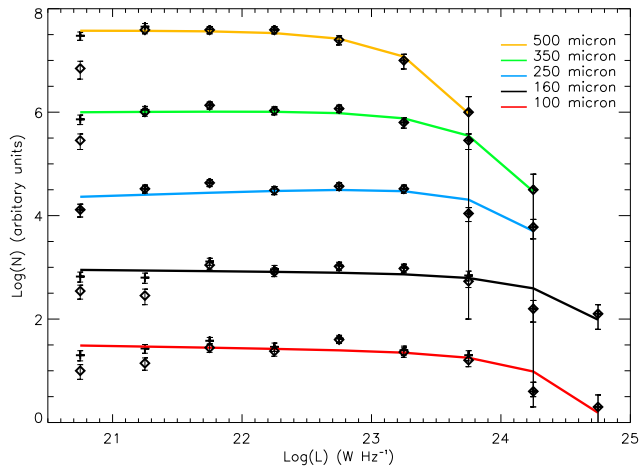}
\caption{Virgo cluster galaxy luminosity functions for the 5 Herschel far-infrared bands. The solid lines are the best fit Schechter functions to the complete adjusted data. The points marked by a diamond are for the unadjusted data. Data has been arbitrarily off-set from each other.}
\end{figure}

The important question is whether there is a faint galaxy excess in the Virgo field that might be associated with a far-infrared population not detected using our optical source list. Looking at Fig. 3 we can see that below about 1 Jy the black line traces the red line very well and there is no evidence for an excess population over and above that detected by our optical selection. To be sure of this conclusion we have assessed the effect of changing our detection threshold. As stated above the variation in detection threshold (calculated within the mesh size) is quite consistent across the NGP field, but does vary across the Virgo field. We have calculated the standard deviation of the 1$\sigma$ fluctuations over fifty 100 sq arc min areas in the Virgo field to see how this influences the Virgo counts. We have used this standard deviation to see how changes in the threshold across the field influence the counts. The effects of these threshold variations is show by the red dashed lines on Fig. 3. Clearly the effect of these changes in the threshold value do not alter our conclusion that the optical selection is picking up most, if not all, of the far-infrared sources in the cluster.

Given the above discussion we conclude that there is no good evidence for an additional population of faint far-infrared sources that is not associated with the previously identified optical sources. Based on this, below we proceed to construct luminosity functions.

\begin{figure*}
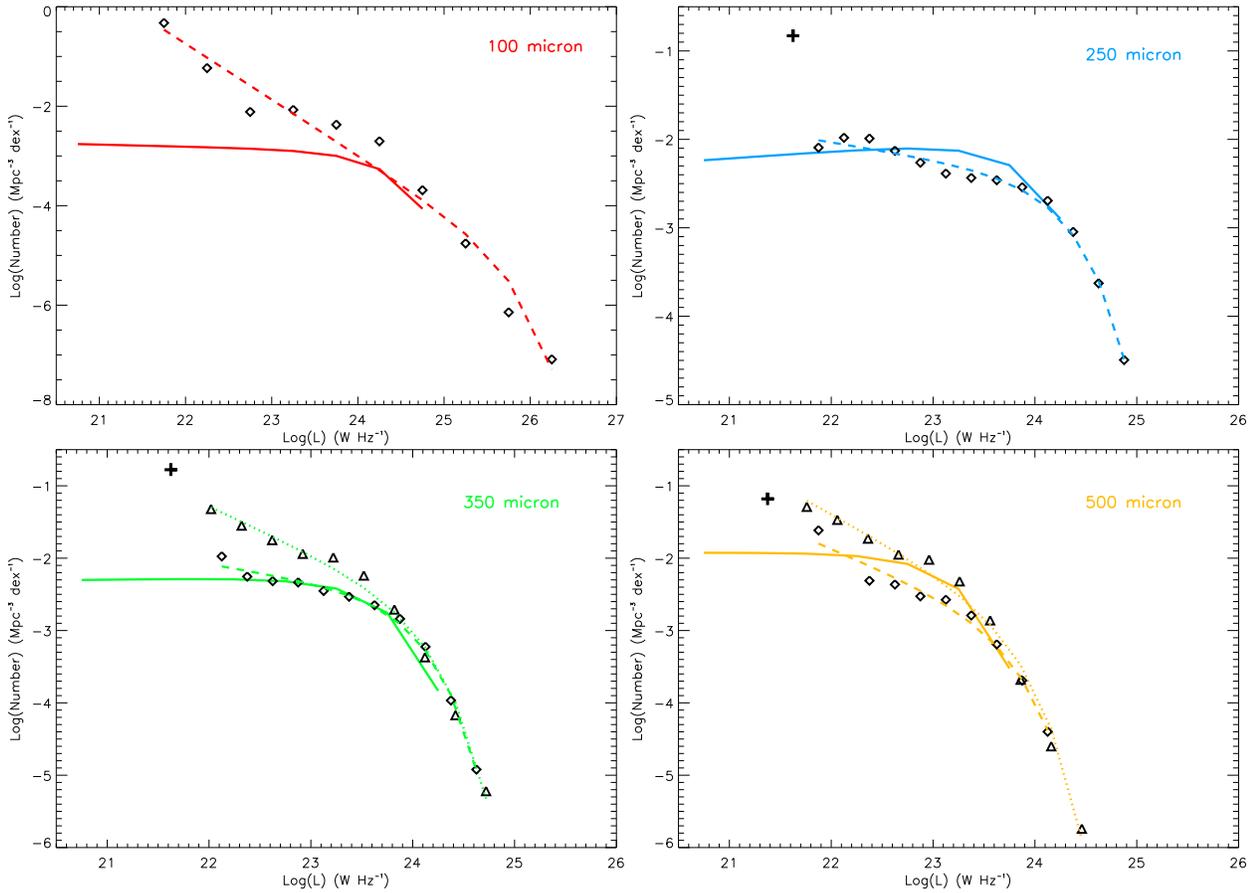

\centering
\includegraphics[scale=0.5]{comparison_100.epsi}
\includegraphics[scale=0.5]{comparison_250.epsi}
\includegraphics[scale=0.5]{comparison_350.epsi}
\includegraphics[scale=0.5]{comparison_500.epsi}
\caption{A comparison of luminosity functions at different wavelengths and over different environments. 100 micron - data points are from the IRAS survey of Sanders et al. (2003) and the dashed line is the Schechter function fit. The solid line is the Virgo luminosity function from Fig. 4 normalised at $L^{*}$. 250 micron - data points come from Eales et al. (2013) and the dashed line is the Schechter function fit ignoring the lowest luminosity point marked by a cross. The solid line is the Virgo luminosity function from Fig. 4 normalised at $L^{*}$. 350 micron - the diamond data points come from Eales et al. (2013) and the dashed line is the Schechter function fit ignoring the lowest luminosity point marked by a cross. The triangular data points come from Negrello et al. (2013) and the dotted line is the Schechter function fit. The solid line is the Virgo luminosity function from Fig. 4 normalised at $L^{*}$. 500 micron - the diamond data points come from Eales et al. (2013) and the dashed line is the Schechter function fit ignoring the lowest luminosity point marked by a cross. The triangular data points come from Negrello et al. (2013) and the dotted line is the Schechter function fit. The solid line is the Virgo luminosity function from Fig. 4 normalised at $L^{*}$.}
\end{figure*}

Using the Auld et al. (2013) data we can greatly extend and update the far-infrared luminosity functions presented in Davies et al. (2012). From the sample definition in Auld et al. (2013) all galaxies have a detection at 250$\mu$m, but not necessarily at the other wavelengths. To make the data sets complete for all 207 galaxies across all wavelengths we have used the mean observed flux density ratios (with 250$\mu$m) to predict the missing flux density values. Of the 207 galaxies 46, 37, 22, 47 galaxies have had their data predicted by this means using ratios ($F_{\lambda}/F_{250}$) of 1.11, 1.34, 0.52 and 0.25 at 100, 160, 350 and 500$\mu$m respectively. 

In Fig. 4 we show the derived luminosity functions with the best fitting Schechter functions. The diamond symbols show the data before the adjustment for missing values was made, the crosses after correcting for missing values. The adjustment predominantly affects the faint end of each luminosity function. The Schechter parameters of each fit are given in Table 1 using where necessary the cluster volume of 62.4 Mpc$^{-1}$. As we will see when comparing to other far infrared luminosity functions the Virgo luminosity functions are characterised by a flat faint-end slope ($\alpha \approx -1.0$). 

To illustrate the disparity between the cluster and field luminosity functions, we also show in Table 1 the Schechter fitting parameters for luminosity functions derived by others - see also Fig. 5. The IRAS 'field' 100$\mu$m data comes from the compilation of 629 galaxies in the Bright Galaxy sample of Sanders et al. (2003). We have simply used the galaxy distances to measure the relevant detection volumes. The Herschel data comes from the H-ATLAS survey (Eales, private communication) using data from the three SPIRE bands at 250, 350 and 500$\mu$m. The Planck data is taken directly from Negrello et al. (2013)
\footnote{We have multiplied the Planck 550$\mu$m flux densities by 1.35 to make them approximately equivalent to the Herschel 500$\mu$m data, see Baes et al. (2013).}.

The most disparate wavelength between cluster and field is 100$\mu$m, where the IRAS luminosity function of nearby bright galaxies is considerably steeper at the faint end than that in the cluster. Sanders et al. (2003) (using the 'bolometric' infrared luminosity rather than the 100$\mu$m flux density used here) actually fit the luminosity function with two power laws, one of slope $\alpha=-0.6$ and at the faint end with $\alpha=-2.2$\footnote{This IRAS bolometric luminosity function is consistent with recent measurements made by Magnelli et al. (2013) of the 8-1000$\mu$m luminosity function based on PACS 70, 100 and 160$\mu$m data.}, compared to our single Schechter function fit with $\alpha=-2.1$. With a slope of $\alpha=-1.0$ our cluster luminosity function either lacks faint dusty galaxies and/or the star formation required to heat the dust. We will return to the issue of dust mass and dust heating in section 4. Another noticeable difference between the cluster and the field is the lack of very luminous infrared sources in the cluster - the derived $L^{*}_{100}$ for the field is about a factor of 20 higher than in the cluster. Given that the density of $L^{*}_{100}$ galaxies in the field is only about $10^{-5}$ Mpc$^{-3}$ even with a cluster over density of about 100 we would still only expect 1 in every 1000 Mpc$^{-3}$ or about a 10\% chance of finding one in our cluster volume. Without studying more clusters we cannot say whether we are unlucky or that clusters do not contain bright far infrared sources, though the lack of bright sources in clusters has previously been noted by Bicay and Giovanelli (1987). With a faint-end slope of $\alpha>-2.0$ the IRAS luminosity function is unbound and we cannot calculate a luminosity density ($\rho_{FIR}=\phi L^{*} \Gamma(2.0+\alpha$)). 

Comparing our Virgo longer wavelength luminosity functions with others, we find that a steeper faint-end slope and a larger value of $L^{*}$ are a common feature of the field. Recently Eales et al. (private communication) have derived the 250, 350 and 500$\mu$m luminosity functions using H-ATLAS data for galaxies with $z \le 0.1$ ($D \le 411$ Mpc). Negrello et al. (2013) have done a similar thing using Planck 350 and 550$\mu$m
data for galaxies with $D \le 100$ Mpc, see Table 1. and Fig.5. When comparing the two the Planck data gives a steeper faint-end slope, about the same $L^{*}$ and a luminosity density a factor of about 2 higher than the Herschel data. Generally the cluster has a far-infrared luminosity density about two orders of magnitude higher than that of the field. The far-infrared luminosity density values given in Table 1 are about a factor of 2 higher than those given in Table 2 of Davies et al. (2012), which were derived by summing the contributions of the bright galaxies rather than fitting and then integrating a Schechter luminosity function.

The issue of the differences between the local 'field' luminosity functions of Herschel H-ATLAS and Planck does not concern us here, but the differences between cluster and field do. It is clear that at all far-infrared wavelengths there is a lack of fainter sources in the cluster compared to what is generally found in the local field. Explanations could be that there is either a relative lack of emitting dust, it is cold or if far-infrared emission is closely connected to star formation (see below) then there is a lack of star formation in low luminosity cluster systems.

\begin{table*}
\begin{center}
\begin{tabular}{c|cccccc}
Band      & Instrument & Region & $\alpha$ & $L^{*}$                 &       $\phi$ & $\rho_{FIR}$ \\
($\mu$m)  &            &        &          & ($10^{24}$ W Hz$^{-1}$) & (Mpc$^{-3}$ dex$^{-1}$)  & ($10^{23}$ W Hz$^{-1}$ Mpc$^{-3}$) \\ \hline
100 & Herschel & Virgo & $-1.0\pm0.1$  & $2.1\pm0.6$ & $0.3\pm0.1$ & 3.3 \\
160 & Herschel & Virgo & $-1.0\pm0.1$  & $2.8\pm1.0$ & $0.3\pm0.1$ & 4.5 \\
250 & Herschel & Virgo & $-0.9\pm0.1$  & $0.8\pm0.2$ & $0.6\pm0.1$ & 2.2  \\
350 & Herschel & Virgo & $-1.0\pm0.1$  & $0.5\pm0.1$ & $0.5\pm0.1$ & 1.2  \\
500 & Herschel & Virgo & $-1.0\pm0.1$  & $0.2\pm0.01$ & $0.5\pm0.1$ & 0.4 \\
100 & IRAS     & Field & $-2.1\pm0.1$  & $46\pm15$    & $0.000012\pm0.000007$ & - \\
250 & Herschel & Field & $-1.19\pm0.04$ & $1.6\pm0.1$   & $0.0017\pm0.0002$ & 0.03  \\
350 & Herschel & Field & $-1.22\pm0.05$  & $0.7\pm0.1$    & $0.0014\pm0.0002$ & 0.01 \\
500 & Herschel & Field & $-1.58\pm0.12$  & $0.4\pm0.1$    & $0.0067\pm0.0003$ & 0.01 \\
350 & Planck   & Field & $-1.65\pm0.08$  & $0.9\pm0.1$    & $0.0013\pm0.0003$ & 0.03  \\
550 & Planck   & Field & $-1.78\pm0.1$  & $0.4\pm0.1$    & $0.0010\pm0.0003$  & 0.02 \\
\end{tabular}
\caption{Schechter function fitting parameters for our Herschel data of the Virgo cluster. For comparisons we also show the fitted parameters for field galaxies using the IRAS 100$\mu$m data of Sanders et al. (2003), the Herschel 250, 350 and 500$\mu$m data of Eales et al. (2013) and the Planck 350 and 550$\mu$m data of Negrello et al. (2013). Note the Planck 550$\mu$m data has been adjusted to correspond to the Herschel 500$\mu$m band - see text.}
\end{center}
\end{table*}

\section{Dust mass, temperature and emissivity index}
We have used the 100-500$\mu$m data to fit modified blackbody curves to the far-infrared spectral energy distributions of the 207 Virgo galaxies detected by Herschel. We have done this in two ways. Firstly, we use a power law dust emissivity $\kappa_{\lambda}=\kappa_{0}(\lambda_{0}/\lambda)^{\beta}$ with $\kappa_{0}=0.192$ m$^{2}$ kg$^{-1}$ at $\lambda_{0}=$350$\mu$m and a fixed $\beta$(=2). Secondly the same as above but now with a variable $\beta$. The fit is obtained using a standard $\chi^{2}$ minimisation technique as is fully described in Smith et al. (2012). As discussed in Davies et al. (2012) (see their Fig. 6) many galaxy far-infrared spectral energy distributions fit modified blackbodies with $\beta=2$ very well, but here we wish to see if we can learn a little more by letting $\beta$ vary as well.

Given the Herschel calibration (Davies et al. 2012) and the fitting procedure (Smith et al. 2012a) we estimate dust mass errors of order 25\% and temperature errors of order 10\% for both methods of fitting. In  Fig. 6 we compare dust masses derived using both methods. In general fixing $\beta=2$ leads to higher dust masses than allowing $\beta$ to be a variable parameter - median dust mass for $\beta=2$ is $3.1 \times 10^{6}$ M$_{\odot}$ while for a free $\beta$ it is about 22\% smaller (see also Bendo et al. 2003, Galametz et al. 2012)\footnote{Note that Bianchi (2013) has shown that these differences in dust mass, calculated for different values of $\beta$ are actually spurious because they are based on a $\beta=2.0$ normalisation ($\kappa_{0}$). So the differences in mass we find are actually a reflection of our lack of knowledge of the dust emissivity function. Throughout the rest of this paper we will use the dust masses calculated using $\beta=2.0$.}. This median dust mass compares with a recent determination of a median dust mass of $5.0 \times 10^{6}$ M$_{\odot}$ for galaxies in the local volume ($d<100$ Mpc) as derived by Clemens et al. (2013) using the Negrello et al. (2013) sample described earlier (same $\beta$, but adjusted by a factor of 1.2 to account for the different emissivity normalisations used). There is no significant difference between the average dust mass of Virgo cluster galaxies and those found in the local field.

Similar differences occur in the derived temperatures. A fixed $\beta$ leads to a median temperature of $18.9$K with a free $\beta$ temperature about 16\% higher, Fig 7. This again compares with a median temperature for galaxies in the Clemens et al. (2013) sample of 17.7K. These temperatures for individual galaxies also compare very well with those recently measured for individually resolved regions in other nearby galaxies (Bendo et al. 2012, Galametz et al. 2012, Smith et al. 2012, Draine et al. 2013) and with the Milky Way equilibrium dust temperature of 17.5K (Lagache et al. 1998). 

\begin{figure}
\centering
\includegraphics[scale=0.52]{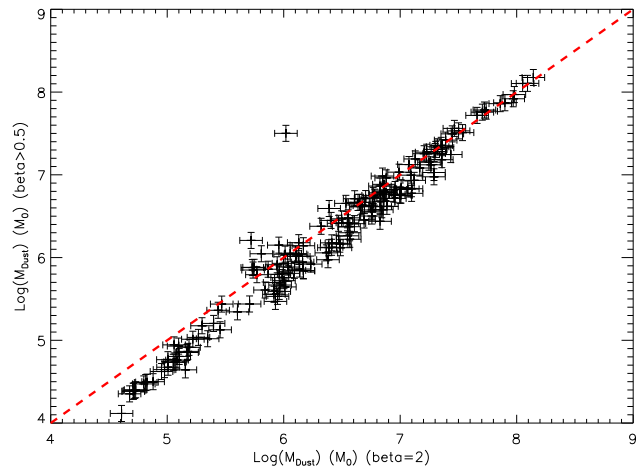}
\caption{The relationship between dust mass derived using a fixed value of dust emissivity index $\beta=2$ and a variable value. The red dashed line is the one-to-one relationship.}
\end{figure}

\begin{figure}
\centering
\includegraphics[scale=0.52]{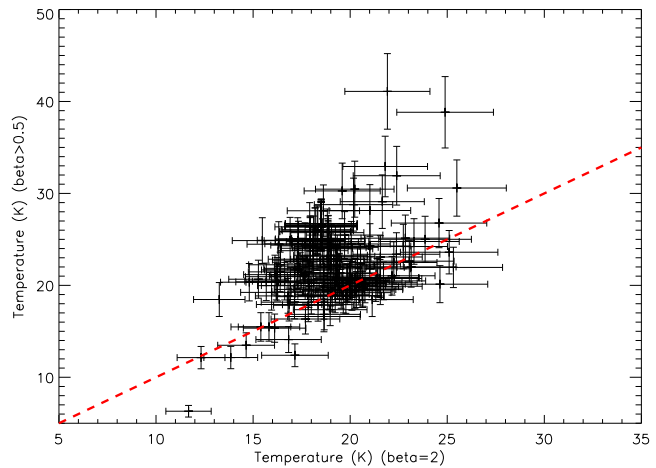}
\caption{The relationship between dust temperature derived using a fixed value of dust emissivity index $\beta=2$ and a variable value. The red dashed line is the one-to-one relationship.}
\end{figure}

One might hope that by deriving individual values for $\beta$ you could learn more about how the properties of the dust might vary between galaxies because the dust emissivity index $\beta$ is related to the physical properties of the dust grains i.e. composition, size and temperature (Skibba et al. 2011, Dunne et al. 2011, Smith et al. 2012a). Low values of $\beta$ are thought to be associated with freshly formed dust in circumstellar disks or stellar winds. $\beta \approx 1$ is thought to be associated with small grains while $\beta \approx 2$ with larger grains formed through grain coagulation or the growth of ice mantels (Seki and Yamamota 1980, Aannestad 1975, Lis etal. 1998, Stepnik et al. 2003). Values for $\beta$ in our Galaxy and in M31 range typically from about 0.5 to 3.0 (Planck Collaboration 2011, Smith et al. 2012a). A similar range for $\beta$ has also been measured by Clemens et al. (2013) using the Planck local volume sample of 234 galaxies (median value of $\beta=1.83$).

193 galaxies in our sample were fitted with the variable $\beta$ model (for 14 galaxies the fit did not converge). For those galaxies with $\beta>0.5$ (see below) the median value of $\beta$ is 1.43 (error on the mean value is 0.06), a little lower than that found by Clemens et al. (2013). In Fig. 8 we plot the derived values of $\beta$ against the derived temperature. Clearly values of $\beta$ fall to lower values ($\beta <0.5$) than have been observed in our Galaxy and M31.  After inspecting the spectra we suggest three reasons why this is so. Firstly, we have in our sample galaxies that not only have thermal emission from dust at these wavelengths, but also to varying degrees contaminating synchrotron emission from an active nucleus. For example the emission from the giant central elliptical galaxy M87 is dominated by synchrotron (Baes et al. 2010, Davies et al. 2012) and when fitted to a modified blackbody gives a spurious $\beta=-2.7$. Secondly, some objects seem to have 'excess' 100$\mu$m emission, which we interpret as being due to a prominent hot dust component associated with star formation (Bendo et al. 2012). In this case a single dust component is again an inadequate interpretation of the data. Thirdly, for some galaxies there is emission at longer wavelengths over and above normal expectations (as observed for example in the Large Magellanic Cloud, Gordon et al. 2010). This additional long wavelength emission can be particularly pronounced in dwarf star forming galaxies (Dale et al. 2012, Remy-Ruyer et al. 2013) and can lead to unrealistically low values of $\beta$.  The latter case leads to what we might describe as 'reasonable values' of the temperature while the first two produce very high unrealistic temperatures. When fitting to a single temperature modified black body values of $\beta<1$ are generally unphysical (Li, 2004). The above highlights the problem of relating $\beta$ to the changing physical properties of the dust.

Given the observation by Smith et al. (2012a), that what we expect to be pure thermal emission from the disc of M31 has $\beta>0.5$, we will use this as a discriminating value for those galaxies that have far-infrared spectral energy distributions that are well fitted by a single modified blackbody - 165 of the 207 galaxies in the sample have $\beta>0.5$. In Fig. 8 we show the relationship between the derived temperature and the emissivity index $\beta$ (those above the yellow dashed line have $\beta>0.5$). There is clearly a relationship between these two parameters. 

The origin of this relation has previously been debated and extensively discussed in Smith et al. (2012a). In summary, it has been proposed that the relationship may not be physical but rather the result of the fitting process (Shetty et al. 2009a) and/or of fitting a one-component modified blackbody to a range of temperatures (Shetty et al. 2009b). This point h,as also been made by Desert et al. (2008) who clearly show how errors in their derived values of $\beta$ and $T$ can lead to a false correlation, but they also show that the range of their data is greater than that expected solely due to these errors. Hence, they suggest that there might be something to be learned about the physical properties of the dust from the relation. As a further indicator that the $\beta-T$ relation has some physical basis Smith et al. (2012a) show that for M31 the relationship is different for regions within and outside a radius of 3.1 kpc, this is the distance from the nucleus of the molecular (dusty) ring. They consider this to be an indicator of different dust within these two regions. If the relationship is physical then it contains important information about the properties of the grains and provides motivation for a grain model that reproduces the relation (Meny et al. 2007, Coupeaud et al. 2011). 

We can compare the $\beta-T$ relation we find with that obtained by others. Included on Fig. 8 are the two best fit lines of Smith et al. (2012a) to the inner (blue) and outer (red) dust of M31 (their equation 6).  Using the same fitting parameterisation ($\beta=A \left( \frac{T}{20K} \right)^{\alpha}$) we find best fitting values of $A=1.58$ and $\alpha=-1.44$ for galaxies with $\beta>0.5$ and types later than S0 (our types 2, 3 and 4, see below). This fits the $R>3.1$ kpc data for M31 almost exactly (see Fig. 8). If the $\beta-T$ relation is physical then the range of conditions found in the disc of M31 are also to be found globally in the different late type galaxies that make up the Virgo sample. With regard to our Galaxy, Desert et al. (2009) have fitted spectral energy distributions to data from the balloon-borne instrument Archeops. They find a similar though flatter $\beta-T$ relation to ours (their value of $T$ for $\beta=2$ is 14.1 while ours is 17K), though this is for individual sources rather than the diffuse emission across the sky, which may be more closely related to the global measurements of galaxies that we have. Clemens et al. (2013) also find a similar $\beta-T$ relation for their local Planck sample and we have plotted this as the large black crosses on Fig. 8 - it is in good agreement with what we find given the temperature errors on their points and our line of typically 2-5K.

\begin{figure}
\centering
\includegraphics[scale=0.52]{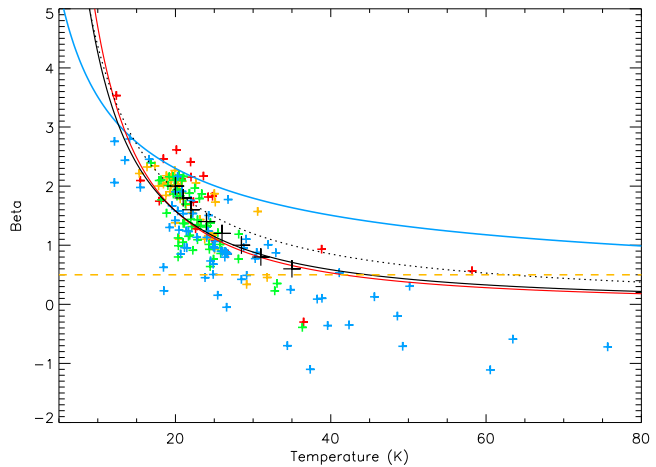}
\caption{The relationship between dust temperature and dust emissivity index for 193 Virgo cluster galaxies. Galaxies are distinguished by their morphological type - earlier than Sa (red), Sa/Sb (yellow), Sc (green) and later than Sc (dwarfs and blue compact dwarfs) (blue). The solid black line is the best fitting line to the model described in the text using 146 galaxies later than S0. The black dotted line is the best fitting line to the model described in the text using 19 galaxies earlier than Sa (red points). The blue line is the relationship derived for regions within $R=3.1$ kpc and the red line for $R>3.1$ kpc as measured for M31 by Smith et al. (2012). The dashed yellow line indicates a value of $\beta=0.5$. The large black crosses indicate the $\beta-T$ relation derived for local field galaxies by Clemens et al. (2013).}
\end{figure}

To look for different dust in different types of galaxies in Fig. 8 we have also identified morphological types (obtained from GOLDMINE) - 21 galaxies earlier than Sa (red), 41 galaxies Sa/Sb (yellow), 54 galaxies Sc (green) and 77 galaxies later than Sc (Sd, dwarfs and blue compact dwarfs) (blue). Roughly Sa/Sb/Sc galaxies occupy the same region of the $\beta-T$ plane as the individual resolved regions of M31 (Smith et al. 2012a). Early type galaxies tend to have higher values of $\beta$ for a given temperature than is typical for the later types (Smith et al. 2012). There is a 'hint' in the data that maybe early type galaxies follow the inner region relationship for M31 (blue line, Fig. 8) rather than the spiral galaxy relationship. The black dotted line is a fit to the early type galaxy data and although it is displaced towards the blue line it is by no means a close fit to it. The greatest scatter in Fig. 8 is produced by those galaxies listed as Sd/dwarf/BCD. As stated above these late type galaxies with low values of $\beta<0.5$ are inadequately fitted by a single modified blackbody because they appear to have a quite prominent warm dust component and so necessitate a more complex SED modelling approach - something we will explore in a later paper.

\begin{figure}
\centering
\includegraphics[scale=0.52]{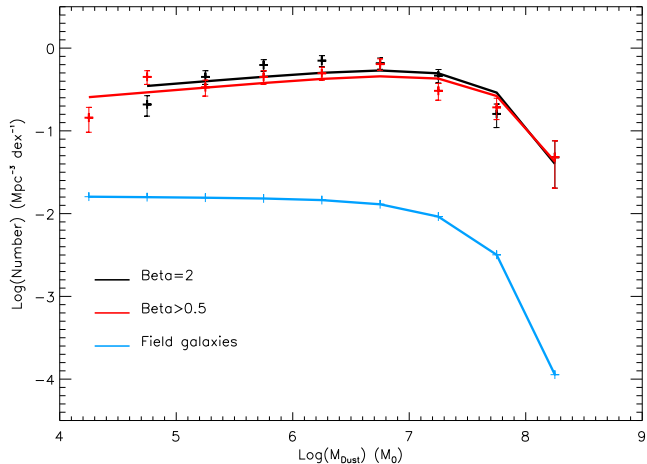}
\caption{The dust mass function. The black line is for all 207 galaxies which have either measured or predicted flux densities in all five Herschel bands and have then been fitted with a $\beta=2$ model. The red line is for the 165 galaxies that have been fitted to a variable $\beta$ model and have values of $\beta>0.5$. The blue line is the field galaxy dust mass function taken from Dunne et al. (2011).}
\end{figure}

Having dust masses for our galaxies we can construct the dust mass function in a similar way to the luminosity functions described earlier (Fig. 9). We have done this separately for the complete sample of 207 galaxies using both the observed and predicted Herschel flux densities and for the sample of 165 galaxies with variable $\beta>0.5$. The fitted mass function parameters are given in Table 2 - they are very similar for both samples with the total dust mass density in the smaller $\beta>0.5$ sample being only about 6\% less than that in the full sample of 207 galaxies. Note that the dust mass function value of M$^{*}_{Dust}$ is very close to the value recently measured for M31 ($5.4 \times 10^{7}$ M$_{\odot}$) by Draine et al. (2013). By using this larger sample and also by fitting a mass function, rather than just summing the contribution of the bright galaxies, we have increased our estimate of the cluster dust mass density, given in Davies et al. (2012), by about a factor of 7 (Table 2). 

We have also compared our dust mass function parameters with those obtained for the general field by Dunne et al. (2011), Table 2. We have used the Dunne et al. (2011) dust mass function because it is derived in a similar way to ours using Herschel data over wavelengths of less than 500$\mu$m. However, we note that the recent dust mass function derived by Clemens et al. (2013) shows a steeper low mass slope of $\alpha=-1.3$, although it compares reasonably well with Dunne et al. at the high mass end ($M_{Dust}>10^{8}$ M$_{\odot}$). The Clemens et al. (2013) dust mass function uses Planck data at wavelengths greater than 500$\mu$m and so this may indicate that even the Herschel data misses cold dust, particularly in lower dust mass galaxies. We are in the process of compiling, where possible, Planck data on the galaxies in this sample (Baes et al., in preparation) .

Continuing our comparison with the Dunne et al. (2011) dust mass function, the derived dust mass density is about a factor of 100 higher in the cluster than it is in the field. All three mass functions shown in Fig. 9 are essentially flat at the low mass end with no evidence for them being different in this respect between cluster and field. This is in contrast to the luminosity functions which were all steeper in the field than in the cluster. This can only come about if we have generally hotter dust in the lower luminosity field galaxies. We might have expected that if dust stripping processes are important in the cluster environment that the relative numbers of low and high dust mass galaxies may have changed between cluster and field i.e. lower mass galaxies more readily losing their dust, but this does not appear to be so. 

\begin{table*}
\begin{center}
\begin{tabular}{c|cccccc}
Sample    & $\alpha$ & $M^{*}_{Dust}$    & $\phi$                    & $\rho_{Dust}$ \\
          &          & ($10^{7}$ $M_{\odot}$) & (Mpc$^{-3}$ dex$^{-1}$) & ($10^{7}$ $M_{\odot}$ Mpc$^{-3}$) \\ \hline
 $\beta=2$   & $-0.9\pm0.1$  & $5.7\pm1.3$ & $0.7\pm0.1$ & 1.8 \\
 $\beta>0.5$ & $-0.9\pm0.1$  & $6.1\pm1.3$ & $0.6\pm0.1$ & 1.7 \\
 Field       & $-1.0$        & $3.6$       & $0.006$     & 0.02  \\
\end{tabular}
\caption{Schechter function fitting parameters to the Virgo cluster dust mass function. For comparisons we also give the values for the field taken from Dunne et al. 2011.}
\end{center}
\end{table*}

\section{Stellar mass}
Where possible we have followed the prescription of Bell et al. (2003) to derive stellar masses using: \\
\begin{center}
$\log(M_{Star})=-0.339 + 0.266(g-r)+\log\left(\frac{L_{H}}{L_{\odot}}\right)$. \\
\end{center}
Optical and near-infrared data were obtained directly from NED for 130 of the 207 galaxies. Where there was insufficient data we have simply used the linear relation between the mass derived using the B band magnitude ($M_{B}^{\odot}=5.48$) only and that calculated using the Bell et al. (2003) formula to estimate stellar masses (blue dashed line Fig. 10). The stellar mass function derived in this way for the 207 galaxies in our Herschel sample is shown as the black line on Fig. 11. The mass function fitting parameters are given in Table 3. The low-mass slope ($\alpha$) for both the dust and stellar mass functions is approximately the same again indicating that there is no preferential removal of dust from low mass systems in the cluster environment. Using the derived densities the stars-to-dust mass ratio for these galaxies is $\sim$1000, as might have been expected from observations of a 'typical' galaxy like the Milky Way.

\begin{figure}
\centering
\includegraphics[scale=0.52]{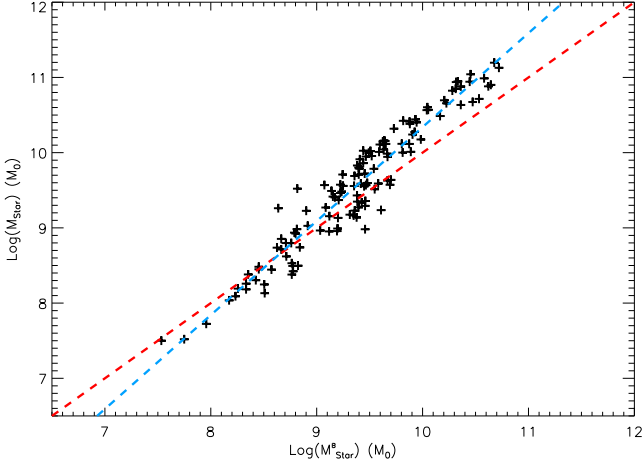}
\caption{The relationship between the stellar masses derived from just B band data ($x$ axis) and that obtained using the method described in the text ($y$ axis). The dashed red line is the one-to-one relationship and the dashed blue line the best fitting line.}
\end{figure}

Of course there are many galaxies detected at optical wavelengths that are not detected by Herschel. Using the same prescriptions as described above we have calculated stellar masses for all 648 VCC galaxies within our survey area with $D=17-23$ Mpc. We have added to this the 43 new galaxies we detected using the SDSS spectroscopic data (section 2) giving them the mean cluster distance of 18.7 Mpc - the VCC+ sample (see appendix 1). The derived mass function is shown as the red line on Fig. 11. The Herschel and VCC+ mass functions coincide at a mass of about $10^{10-11}$ M$_{\odot}$ with there being some optically bright early type galaxies that are not detected by Herschel as well as those that are optically faint. 

How representative the VCC is of the totality of galaxies in the Virgo cluster has been the subject of much debate in the past. Derived faint-end slopes for the optical luminosity function range from about $\alpha=-1.2$ to -2.0 (Impey et al. 1988, Phillipps et al. 1998, Sabatini et al. 2003). The crucial problem is what remains hidden beneath the rather high surface brightness limit imposed by both the optical data used for the VCC and the SDSS spectroscopic data. Hopefully the 'Next Generation Virgo Cluster Survey' (Ferrarese et al. 2012) with its low surface brightness sensitivity will soon pass judgement on this issue. In this paper we will use the stellar mass densities given in Table 3, but note that if all else remains the same then increasing the faint-end slope of the mass function from, -1.2 to -1.7 for example, increases the stellar mass density by a factor of about 2.6. The stellar mass density of the VCC+ sample is about a factor of 1.8 higher than that derived using just the Herschel galaxies. This all leads to a cluster stars-to-dust ratio of about 1800 - almost twice the value measured for individual galaxies in the Herschel sample - there are of course many galaxies with stars, but no detectable dust emission. 

For comparison also included on Fig. 11, with fit parameters listed in Table 3, is the field galaxy stellar mass function of Panter et al. (2007). The comparative faint-end slope of the luminosity function in clusters and in the field has been an issue of much debate (Phillipps et al. 1998, Roberts et al. 2004), but is not a problem for us here other than to note that both the VCC+ and Field galaxy samples lead to stellar mass functions with the same low mass slope of $\alpha=-1.2$. The stars-to-dust ratio for the field (Tables 3 and 4) is about 1500 reasonably consistent with that derived above for the cluster (1800). This yet again indicates that these cluster galaxies have not preferably lost dust when compared to galaxies in the field. We again note the lack of galaxies in the cluster with large stellar masses, but as stated above this may just be due to the rarity of bright galaxies and the relatively small volume we are considering.

\begin{figure}
\centering
\includegraphics[scale=0.52]{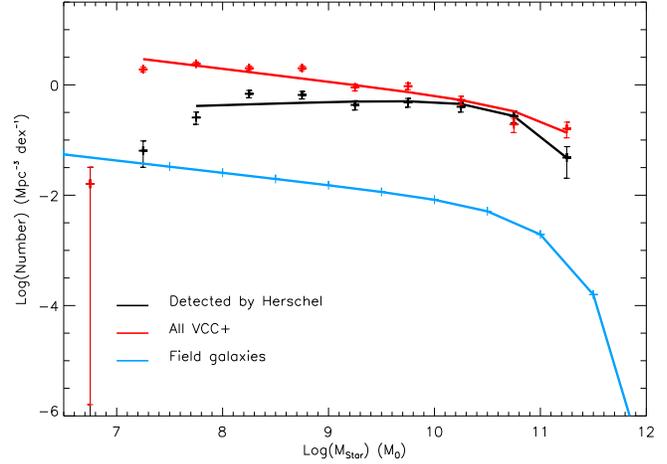}
\caption{The stellar mass function. The black line and data points are for all 207 galaxies in the Herschel sample.  The red line and points are for the 691 VCC+ galaxies in our survey area that have a distance of 17 to 23 Mpc. The blue line is the field galaxy stellar mass function taken from Panter et al. (2011).}
\end{figure}

\begin{figure}
\centering
\includegraphics[scale=0.52]{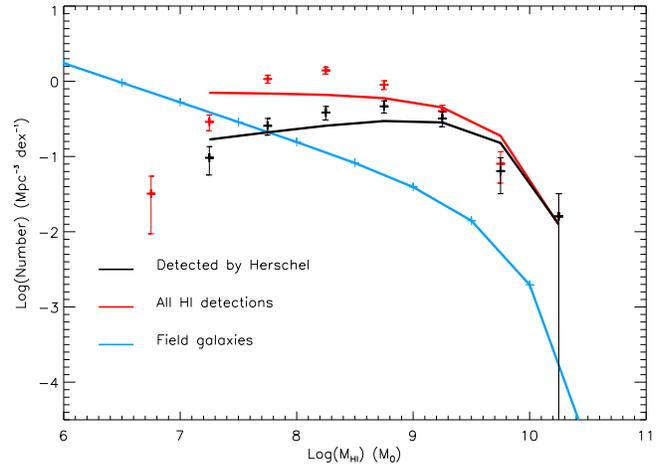}
\caption{The HI mass function. The red line and points are for the 261 galaxies within our Herschel fields that are detected by the ALFALFA survey. The black line and data points are for 133 galaxies in the Herschel sample that also have a HI detection. The blue line is the field galaxy HI mass function taken from Davies et al. (2011).}
\end{figure}

\begin{table*}
\begin{center}
\begin{tabular}{c|cccccc}
Sample    & $\alpha$ & $M^{*}_{Stars}$    & $\phi$                    & $\rho_{Stars}$ \\
          &          & ($10^{10}$ $M_{\odot}$) & (Mpc$^{-3}$ dex$^{-1}$) & ($10^{10}$ $M_{\odot}$ Mpc$^{-3}$) \\ \hline
 Herschel   & $-0.9\pm0.1$  & $6.6\pm1.4$ & $0.6\pm0.1$ & 1.8 \\
 All VCC & $-1.2\pm0.1$  & $19.2\pm11.7$ & $0.3\pm0.1$ & 3.3 \\
 Field       & $-1.2$        & $9.5$       & $0.002$     & 0.03  \\
\end{tabular}
\caption{Schechter function fitting parameters to the Virgo cluster stellar mass function. For comparisons we also give the values for the field taken from Panter et al. 2007.}
\end{center}
\end{table*}

\section{Atomic gas mass}
We have taken our atomic hydrogen data from the ALFALFA database (Giovannelli et al. 2007). The ALFALFA survey does not completely cover our Herschel area missing declinations below 3.8$^{o}$ and above 16.2$^{o}$. So, we select data from the survey over this range in declination and over a range of 6.8$^{o}$ in right ascension to give us the same area on the sky as our Herschel survey. We have selected all objects in this area that have a velocity of $400<v_{Helio}<2659$ km s$^{-1}$. The upper bound corresponds to the highest velocity in our Herschel detection list of 207 galaxies and the lower bound is set to avoid confusion with Galactic hydrogen and local high velocity clouds. With this area and velocity range we hopefully sample about the same volume as the stellar and dust selected samples. Within this volume there are 261 HI detections in the ALFALFA catalogue, 65 of these are not in the VCC and are listed in appendix 2. We have used the ALFALFA catalogued values for HI mass and distance and plotted the corresponding HI mass function in Fig. 12 (red line), Schechter function fit parameters are given in Table 4. Using just those galaxies in the Herschel sample that have a HI detection we have 133 galaxies, their HI mass function is shown as the black line in Fig. 12, again with Schechter parameters given in Table 4. Finally, on Fig. 12 we show the field galaxy HI mass function (blue line)  derived by Davies et al. (2011), parameters again in Table 4. Note that the HI mass of M31 is just a little higher ($6.6 \times 10^{9}$ M$_{\odot}$) than the values of M$^{*}_{HI}$ we obtain here.

\begin{figure}
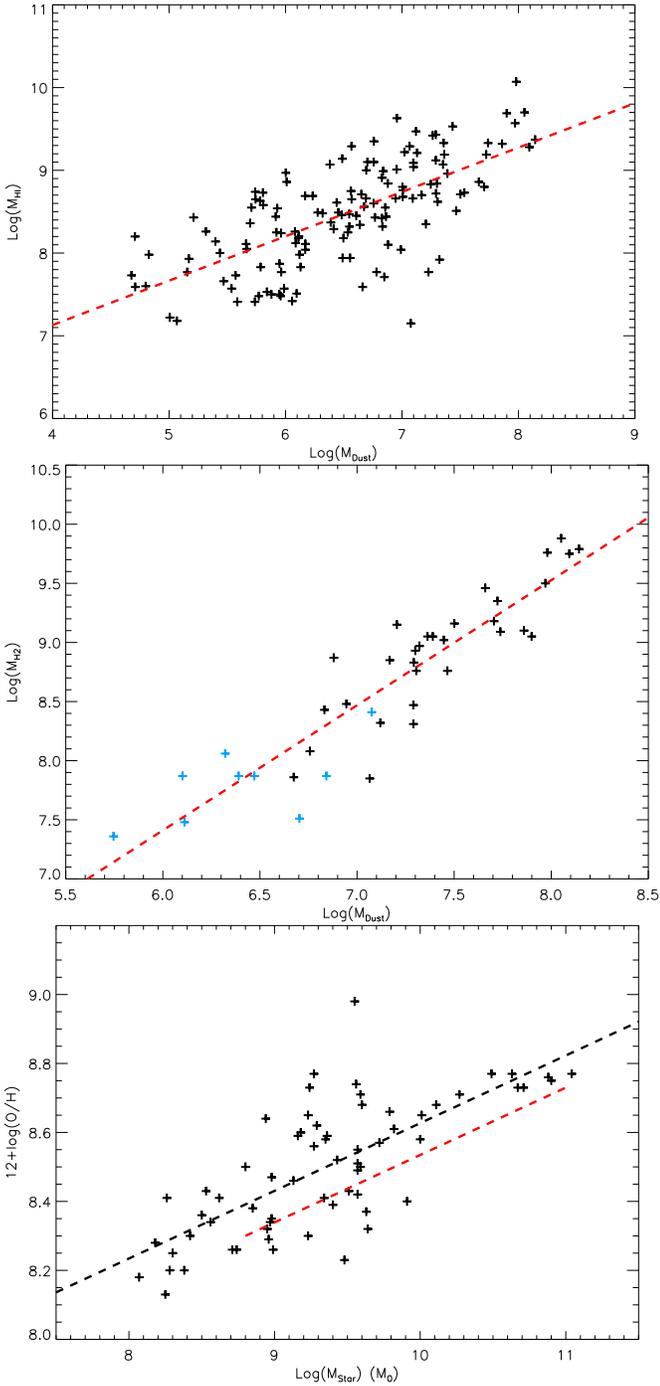

\centering
\includegraphics[scale=0.52]{mass_dust_gas.epsi}
\includegraphics[scale=0.52]{mass_dust_h2.epsi}
\includegraphics[scale=0.52]{mass_metalicity.epsi}
\caption{Top - The relationship between calculated ($\beta=2$) dust mass ($M_{Dust}$) and atomic hydrogen mass ($M_{HI}$) - the red dashed line is the linear least squares fit to the data.  Middle - mass of dust ($M_{Dust}$) against $M_{H_{2}}$. The black points are for 30 late type galaxies from Corbelli et al. (2012) and the blue points for 9 early type galaxies from Young et al. (2011) - the red dashed line is the linear least squares fit to the data. Bottom - the stellar mass ($M_{Stars}$) against the oxygen abundance for our sample galaxies is shown as black crosses with the black dashed line showing the linear least squares fit to the data. The red dashed line is the relationship for non-Virgo cluster galaxies taken from Hughes et al. 2013 (their Fig. 7).} 
\end{figure}

Whereas there is little indication that the dust and stellar mass of low mass galaxies is greatly affected by the cluster environment, because the cluster and field values of $\alpha$ are effectively the same, this is not true for the atomic hydrogen. The low mass slope of the field galaxy HI mass function is significantly steeper than that of the cluster. The cluster environment is certainly affecting the HI content of low mass galaxies. For example, from Table 2 the ratio of dust mass densities between cluster and field is $\sim90$, the same ratio for stars from Table 3 is $\sim110$, while that for atomic gas is $\sim20$. If the cluster is an over density in baryons by a factor of about 100, as measured by the stars and dust, then the galaxies are depleted in HI by about a factor of 4. This is not a new result as gas depletion in Virgo spirals is discussed extensively in Haynes et al. (1984).

What is the origin of this depletion? Broadly there are two options either the gas has been lost from the galaxies due to their presence in the cluster, most likely by ram pressure, or the gas has been consumed during star formation. We will consider this issue further in section 8. 

If we are trying to understand the differences in the interstellar medium between galaxies in the cluster and the field, that arise because of the environment, one thing to consider is gas(atomic)-to-dust ratios. In the field this ratio is about $\sim400$ while in the cluster as a whole it is $\sim63$ (see Tables 2 and 4).  

\begin{table*}
\begin{center}
\begin{tabular}{c|cccccc}
Sample    & $\alpha$ & $M^{*}_{HI}$    & $\phi$                    & $\rho_{HI}$ \\
          &          & ($10^{9}$ $M_{\odot}$) & (Mpc$^{-3}$ dex$^{-1}$) & ($10^{9}$ $M_{\odot}$ Mpc$^{-3}$) \\ \hline
 Herschel   & $-0.8\pm0.2$  & $4.1\pm1.2$ & $0.6\pm0.2$ & 1.1 \\
 All ALFALFA & $-1.0\pm0.2$  & $4.5\pm1.6$ & $0.6\pm0.3$ & 1.3 \\
 Field       & $-1.5$        & $5.0$       & $0.009$     & 0.08  \\
\end{tabular}
\caption{Schechter function fitting parameters to the Virgo cluster HI mass function. For comparisons we also give the values for the field taken from Davies et al. 2011.}
\end{center}
\end{table*}

If we are interested in the total baryon budget in galaxies and also if we want to consider the chemical evolution of galaxies then we require not just the mass in atomic hydrogen, but the total mass of gas. This includes molecular hydrogen, helium and the mass in the diffuse ionised warm and hot components. These are issues we will discuss in the next section. 

\section{The baryon budget}
In the above sections we have measured and described three important constituents of galaxies, dust, stars and atomic gas, but this is not the totality of the baryons. Although important the dust and atomic gas components are incomplete because they do not measure all the metals or all of the gas. The dust is only representative of the total amount of metals and the atomic gas of the total gas mass. In this section we will use some quite sparse data and fits to mass functions to make an estimate of the total mass in metals and gas in our Herschel sample galaxies accepting that we have already derived the mass in stars.

As not all of the 207 Herschel sample galaxies have an HI mass, for 74 galaxies we have used the dust HI mass correlation to predict a HI mass, Fig 13 (top). The linear least squares best fitting line is $\log{M_{HI}}=(0.53 \pm0.01) \log{M_{Dust}} + (4.98 \pm0.04)$.

We have obtained from the literature molecular hydrogen gas masses for 39 of our Herschel sample galaxies - 30 late types from Corbelli et al. (2012) and 9 early types from Young et al. (2011). In each case the masses were derived from CO (J=1-0) observations using a constant CO line flux to molecular hydrogen conversion factor of $X_{CO}=2\times10^{20}$ cm$^{-2}$ (K$^{-1}$ km s$^{-1}$)$^{-1}$ (Strong and Mattox, 1996). For the 168 galaxies without a H$_{2}$ mass we have used the dust H$_{2}$ mass correlation to predicted a H$_{2}$ mass, Fig 13 (middle). The linear least squares best fitting line is $\log{M_{H_{2}}}=(1.06 \pm0.02) \log{M_{Dust}} + (1.06 \pm0.12)$. Note that the relation between molecular gas and dust is approximately linear while that between atomic gas and dust goes approximately as the square root of the dust mass. Only 32 galaxies have both a HI and H$_{2}$ mass and for these the mean ratio of atomic to molecular hydrogen is 4.2. Recently for M31 Draine et al. (2013) obtained a value of $M_{HI}/M_{H_{2}}=19.8$, but the HI was measured over 25 kpc and the H$_{2}$ over 12 kpc. Making a simple adjustment for the areas (constant surface density) leads to a value of $M_{HI}/M_{H_{2}}=4.3$ very close to our mean ratio.
We have also corrected the gas mass for the abundance of helium using $M_{H}/M_{He}$=3.0. 

The above provides the correction for cold gas ($T<10^{3}$ K), but there is also the hot/warm diffuse ionised gas component of the interstellar medium, which we know much less about. Much of what we know of this ionised component is derived from observations of the Milky Way, very little is known about its contribution to the baryon content of external galaxies. A comprehensive review of the warm component ($\approx10^{4}$ K) is given in Haffner et al. (2009). They discuss $H_{\alpha}$ observations both of our galaxy and others particularly edge-on galaxies where the diffuse emission can be seen to extend above the mid-plane and above the stars. 
In summary, Haffner et al. say that the diffuse warm component accounts for about 90\% of the ionised hydrogen in galaxies and that this is about one third of the atomic mass. This is confirmed by recent observations of sight-lines to stars in our Galaxy, which again have a mean value of about one third for the ionised to atomic components (Howk and Consiglio, 2012). 

With regard to the hot component ($10^{5}-10^{7}$ K) there have been a number of observations that infer that this may be a substantial fraction of the total gas mass. One of the main issues is whether this hot gas is actually associated with individual galaxies or if it resides within larger filamentary structures (Tripp et al. 2000, Gupta et al. 2012). Gupta et al. (2012) have recently claimed that the hot component of the Milky Way has a mass equivalent to that of the stars, in which case it would be the dominant gas phase component. There are two issues with regard to this: the first is that, as said above this could be material external to the galaxy (particle velocities are of order the Galactic escape velocity). Secondly, the values Gupta et al. (2012) use for the solar oxygen abundance and the metallicity of the hot gas are crucial in their calculation. They use a value for the solar oxygen abundance a factor of 1.75 larger than the more recent determination we will use below and although they say that the metallicity of the hot gas is expected to be about 0.2$z_{\odot}$ they then go on to use 0.3$z_{\odot}$. These two factors combined lead to a factor of 18 increase in the mass derived (they are cubed in the total mass equation). With a factor of 18 decrease in the derived mass the hot component becomes about one half of the atomic component. Our conclusion is that at the moment it is very difficult to accurately account for the baryons that reside in the combined warm and hot components of galaxies. We have conservatively assigned a mass equivalent to the atomic mass to these two components when we have created our total gas mass function. 

For our Herschel sample of 207 galaxies we have used the above correlations and assumptions to obtain total gas masses. For our HI selected sample (ALFALFA) we have used $M_{Gas}^{Tot}\approx2.5M_{HI}$ derived from the HI mass, the mean ratio of molecular to atomic gas, the abundance of helium and the gas in the warm and hot components to get total gas masses.

\begin{figure}
\centering
\includegraphics[scale=0.52]{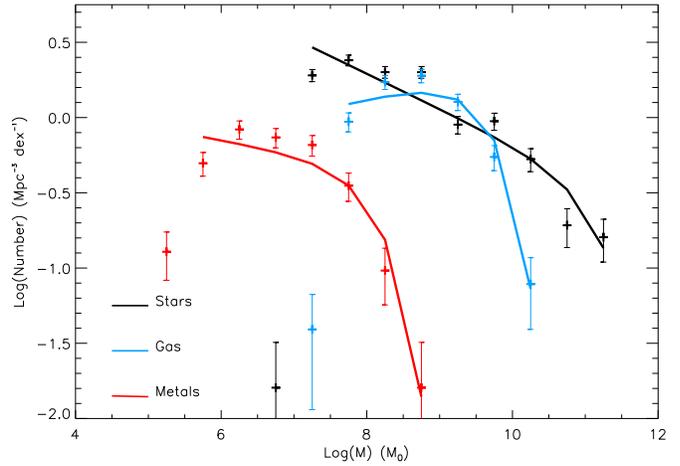}
\caption{The mass functions for the three baryonic components of galaxies in the Virgo cluster. Black total stellar mass, blue total gas mass and red total mass in metals.} 
\end{figure}

We have also obtained from the literature (See Hughes et al. 2013, Table 2) oxygen abundance values for 65 galaxies from our Herschel sample. Oxygen abundances were obtained using drift scan optical spectroscopy and are based on metallicity calibrations from Kewley and Ellison (2008). We have used the stellar mass metallicity relation (Fig 13) to assign a metallicity to those galaxies in the Herschel sample that do not have a measured value - a linear least squares fit gives $12.0+\log{(O/H)}=0.20\pm0.01\log{M_{Stars}}+6.66\pm0.01$. By looking at the residuals after subtracting this linear least squares fit from the data we estimate about a 30\% error on metalicities calculated in this way. To go from the oxygen abundance to the total metallicity ($z$) i.e. the total mass of metals in the gas phase, we require the oxygen abundance of the Sun and its metallicity. Asplund and Garcia-Perez (2001) give the solar oxygen abundance as $12.0+\log{(O/H)_{\odot}}=8.69$ and $z_{\odot}=0.014$ which gives $z=29.2(O/H)$. The total mass of metals in the gas is then just $zM_{HI}$. Using these numbers we find for our sample galaxies $<z>=0.0093 \pm 0.0003$, about two-thirds of the solar value quoted above, and that the mean fraction of metals in the dust is $0.50 \pm 0.02$. The fraction of metals in dust has previously been estimated to be 0.5 by Meyer et al. (1998) and Whittet (1991) and 0.4 by Dwek (1998). Finally, we need to add to this the metals that are in the warm and hot component, which from above we estimate to be $0.2zM_{HI}$ (Gupta et al. 2012) giving $M_{Metals}=1.2zM_{HI}+M_{Dust}$.

Putting all these things together we can create total mass functions for stars, gas and metals residing in Virgo cluster galaxies. The total stellar mass function is just that given in Fig 11 and is derived from the optically selected VCC catalogue with additional galaxies selected via their redshift from SDSS (VCC+, 691 galaxies). The total gas mass function is that derived from the HI selected ALFALFA data with each galaxy's gas mass adjusted for molecular hydrogen, helium and gas in the warm and hot components (261 galaxies). The total metals mass function comes from our Herschel data (selected via their far-infrared emission), which provides dust masses and to this we add the mass of gas phase metals (207 galaxies). 

The mass functions are shown in Fig. 14. and the parameters of the best fitting Schechter function are given in Table 5 - in each case the lowest mass point is omitted from the fit. The bottom line is that cluster values for the ratios $M_{Stars}:M_{Gas}:M_{Metals}$ as a whole (in galaxies) correspond almost exactly with the canonical values that are normally quoted for the Milky Way i.e. that the stellar mass is about 10 (8) times the gas mass and that the mass in gas is about 150 (130) times that in metals - values in brackets are those calculated from Table 5. Although the HI mass function on its own shows that galaxies in the cluster are relatively depleted in atomic hydrogen, the inclusion of helium, molecular, warm and hot gas takes us back to familiar ground.

\begin{table*}
\begin{center}
\begin{tabular}{c|cccccc}
Baryonic    & $\alpha$ & $M^{*}$    & $\phi$                    & $\rho$ \\
Component          &          & ($M_{\odot}$) & (Mpc$^{-3}$ dex$^{-1}$) & ($M_{\odot}$ Mpc$^{-3}$) \\ \hline
 Total stars  & $-1.2\pm0.1$  & $1.9\pm1.2\times10^{11}$ & $0.3\pm0.1$ & $3.3\times10^{10}$ \\
 Total gas    & $-0.9\pm0.1$  & $5.0\pm0.8\times10^{9}$ & $1.8\pm0.3$ & $4.3\times10^{9}$ \\
 Total metals & $-1.1\pm0.1$ & $1.7\pm0.3\times10^{8}$       & $0.4\pm0.1$     & $3.3\times10^{7}$  \\
\end{tabular}
\caption{Schechter function fitting parameters to the Virgo cluster total mass functions. This is the total mass contained within galaxies. The final column gives the mean cluster densities, which of course vary considerably through out the cluster. A 1 Mpc box around M87 contains $13.5\times10^{10}$ $M_{\odot}$ Mpc$^{-3}$ of stars due to M87 alone (Davies et al. 2012).}
\end{center}
\end{table*}

\section{Galaxy scaling relations}
In this section we consider four scaling relations for galaxies. Firstly, the relation between gas fraction and metallicity and its interpretation using a chemical evolution model. Secondly, the relation between stellar mass and the current star formation rate and hence the specific star formation rate of galaxies. Thirdly and fourthly, the baryonic Tully-Fisher relation and the mass size relation, both of whose origin must presumably lie in the gravitational stability of galaxies. Wherever possible we will compare our results for the Virgo cluster with those obtained using galaxies that sample the more general galaxy population.

\subsection{Chemical evolution}
Having the three major baryonic constituents of Virgo cluster galaxies (stars, gas, metals) we can now see if the mass ratios between them are consistent with a chemical evolution model. The simplest model, yet one that provides an insight into how a galaxy evolves chemically, is the closed box model (Edmunds, 1990 and references therein). In its simplest form this model describes the growth of the fractional mass of metals $z$ in the interstellar medium as a function of $p$ the stellar yield and $f$ the gas fraction. $p$ is the fractional mass of metals per unit mass of gas freshly formed in nucleosynthesis. The above parameters are simply related via $z \le p \ln{(1/f)}$. The equality in this equation applies  to the closed box model in which there are no inflows or outflows of gas as the galaxy evolves. More complex models in which various forms of inflow and outflow are described can be found in Edmunds (1990). Edmunds (1990) defines the effective yield as $p_{eff}=\frac{z}{\ln{(1/f)}}$ i.e. the derived yield  irrespective of whether there are inflows or outflows, he also makes some generalised comments on these cases. For example models with outflow, but no inflow or inflow of gas with relatively low metallicity have $p_{eff}<p$. This is straight forward to understand because un-enriched inflow dilutes the interstellar medium while enriched outflow reduces the gas fraction at the same metallicity. Thus the model can provide an insight into how a galaxy in a specific environment has evolved. Using the data for our 207 Herschel galaxies with either measured or predicted values of $M_{Star}$, $M_{Gas}$ and $M_{Metals}$ we can compare our data with this simple closed box model. 

In Fig. 15 we show the derived value of $z_{Tot}=M_{Metals}/M_{Gas}$ plotted against $\ln({1/f})$. \footnote{Note: $z_{Tot}$ is not the same as the metallicity derived from the oxygen abundance. It includes the mass of metals in the dust and the contribution of helium, H$_{2}$, and warm and hot gas to the gas fraction.} Note that the range of metalicities found in individual galaxies is just about the same as that found within different regions of a single galaxy. Within M31 Draine et al. (2013) find a variation of metallicity from about 3$z_{\odot}$ at the centre to about 0.3$z_{\odot}$ in the outer regions.

If the yield $p$ has a constant value for all galaxies i.e. purely determined by the physical processes within stars, then we would expect the data shown in Fig. 15 to lie on a straight line if they evolve as closed boxes, they clearly do not follow this relationship. This result has been known for some time - galaxies do not evolve as closed boxes - however the important issue here is whether cluster galaxies have values of $p_{eff}$ that have been significantly affected by their environment.  To decide on this issues we require a value for the yield $p$. Vila-Costas and Edmunds (1992) give a value for $p$ in the rather wide range 0.004-0.012. On Fig. 15 we have plotted the two lines (dashed blue) defined by these two values and most of our sample galaxies do lie between these two extremes. The mean value for $p$ obtained from our data is $0.009 \pm 0.001$ (dashed black line) consistent with the Vila-Costas and Edmunds (1992) values and with the value of 0.0104 obtained by Tremonti et al. (2004) in their recent chemical evolution model of $\sim53,000$ SDSS galaxies. We conclude that on average Virgo cluster galaxies have a derived value of $p$ consistent with other galaxy samples of predominantly non-cluster galaxies. We find no correlation of $p_{eff}$ with metallicity as might be expected if the fractional mass of metals released back into the interstellar medium is dependent on metallicity, so we assume for the moment that $p$ is a constant and that different positions occupied by galaxies on Fig 15 reflect changes not in $p$, but in $p_{eff}$ because of the in or out flow of gas. Galaxies are clearly segregated in Fig. 15 when it comes to morphology, though we note that the contribution of hot gas may have been underestimated for the early type galaxies, in which case the red data points would move to the left. 

Based on the assumption that $p$ is a constant data points below the black line on Fig. 15, and more convincingly below the lower blue line, have values of $p_{eff}$ lower than might be expected due to stellar processes and could be the result of gas loss. To gain further insight into this issue we have looked at the relation between $p_{eff}$ and other galaxy properties. In Fig. 16 we plot $p_{eff}$ against the total mass in baryons. Previously Tremonti et al. (2004) have found, using $\sim53,000$ SDSS galaxies that $p_{eff}$ increases with baryonic mass - red dashed line Fig. 16. Their interpretation of this result is that lower mass galaxies suffer proportionately more from gas loss and so their effective yield is lower i.e. lower than expected metallicity at a given gas fraction because gas has been lost instead of consumed in stars. Results from the Lee et al. (2003) study of Virgo dwarf irregular galaxies qualitatively support this conclusion (Fig. 16). Within our data we find no correlation between galaxy total mass (baryonic) and $p_{eff}$, but we do need to qualify our result. We derive $p_{eff}$ using the metals in both the gas and the dust, not just those in the gas as used by both Tremonti et al. (2004) and Lee et al. (2003).  We also use our estimate of the total gas mass while Tremonti et al. resort to inferring the gas mass from the star formation gas density relation while Lee et al. use the mass of atomic hydrogen. There may also be a selection effect here because we selected galaxies via their emission from their Interstellar medium (dust) - Tremonti et al. selected galaxies via emission from stars. As our galaxies have to have an interstellar medium to be detected maybe the low mass galaxies in our sample that still have their interstellar medium are young i.e. there as been insufficient time to have as yet undergone gas loss. Those that have undergone gas loss are just not in our sample.

Emphasised again in Fig. 16 is the trend of increasing values of $p_{eff}$ when going from early to late types. However, investigating this further what we find is a clear relationship between galaxy colour and $p_{eff}$ - red galaxies have lower values of $p_{eff}$ - Fig. 17 (top). Given the lack of a mass $p_{eff}$ relation we put this down to an age effect, which fits in with what we said above about a selection effect. The older a galaxy is the more it seems to have been influenced by gas loss processes and $p_{eff}$ has become significantly lower than $p$. As early type galaxies tend to be red we should also see this in a morphology $p_{eff}$ relation, which we clearly do see in Fig. 17 (bottom). This idea of substantial gas loss by early type galaxies has been used to explain the origin of metals in the intra-cluster X-ray gas and will be discussed further in section 9. 

In summary, although the faint end of the HI mass function is flatter than in the field we do not see any evidence in the chemical evolution that low mass galaxies have preferentially suffered from gas loss. The simplest explanation is that mass loss is catastrophic such that we only now see in our sample selected via its interstellar medium those galaxies that are yet to be affected. As the mass loss seems to be a consequence of being in the cluster (HI mass function) then either these galaxies are young and/or they are recent arrivals. Contrary to this it is early type, not necessarily low mass, galaxies that show the most clear cut signs of gas loss.

If individual galaxies do not behave as closed boxes does the cluster as a whole? According to White et al. (1993) the baryon fraction in clusters does not change with time - they act as a closed box retaining all information about past star formation and metal production. If true we can use the stellar, gas and metals mass of the cluster as a whole i.e. the mass densities given in Table 5, which are derived from the integrals of the mass functions, in a chemical evolution model of the cluster. Note that this is for material in galaxies - we will consider material in the intra-cluster medium in the next section. These derived mass densities define the total cluster gas mass fraction and metallicity due to material in galaxies. This data point is plotted on Fig. 15 as a large black star and is consistent with a value of $p_{eff}$ that has been significantly affected by galactic gas mass loss. In this instance what we mean by gas loss is that it is gas lost by the galaxies, so not available for continued star formation, but it is retained within the cluster. 

We have used the outflow model of Dunne et al. (2011) to assess the implications of this gas loss. In their model we now have $g=\frac{f}{1+R(1-f)}$ where $g$ is the gas mass when there is outflow. $R=\lambda/\alpha$ where $\lambda$ is the ratio of gas loss rate to the star formation rate and $\alpha=0.7$ is the fraction of mass from each generation of star formation tied up in long lived stars. Our chemical evolution model now becomes $z=\frac{p\ln(1/g)}{1+R}$. Using a value of $R=2$ leads to the red dot-dash line on Fig. 15 that goes through the data point for the cluster as a whole. A value of $R=2$ implies that 1.4 times the mass of the stars has been lost from the cluster galaxies. If the cluster evolves as a closed box, retaining this material within the cluster environment, then this lost gas must still reside within the intra-cluster medium (section 9).  

\begin{figure*}
\centering
\includegraphics[scale=0.8]{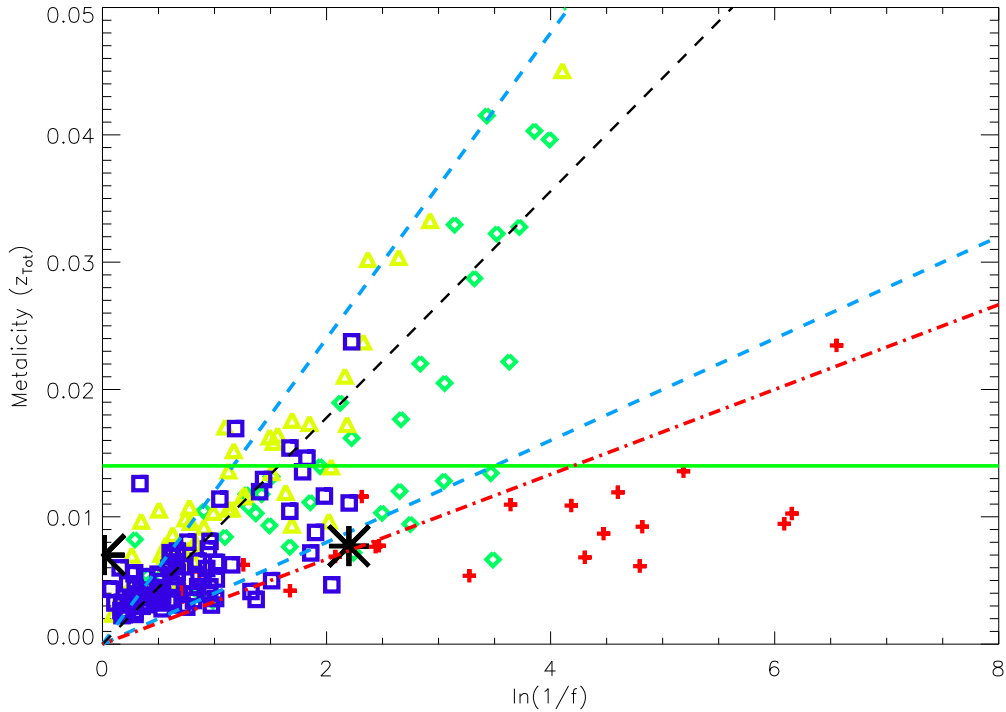}
\caption{The derived metallicity (metals in both dust and the gas phase) against $\ln{(1/f)}$ for our sample of 207 galaxies, $f$ is the gas fraction. It is difficult to estimate errors on $f$ given the assumptions we have made (see text), we expect about 30\% errors on the metalicities. Red crosses are galaxies with type earlier than Sa, green diamonds type Sa/Sb, yellow triangle type Sc and blue square galaxies later than Sc and dwarfs. The blue dashed lines are for a closed box model with a yield of $p=0.004$ (bottom) and $p=0.012$ (top). The solid green line marks the metallicity of the Sun ($z=0.014$). The red dot-dash line is the mass loss model of Dunne et al. (2011) with a mass loss of 1.4 times the star formation rate. The black dashed line is the best fitting line to the data. The large black star to the right marks the data for the cluster galaxies as a whole using the total mass densities given in Table 5. The large black star to the left marks the data for the cluster as a whole including X-ray gas.} 
\end{figure*} 

\begin{figure}
\centering
\includegraphics[scale=0.52]{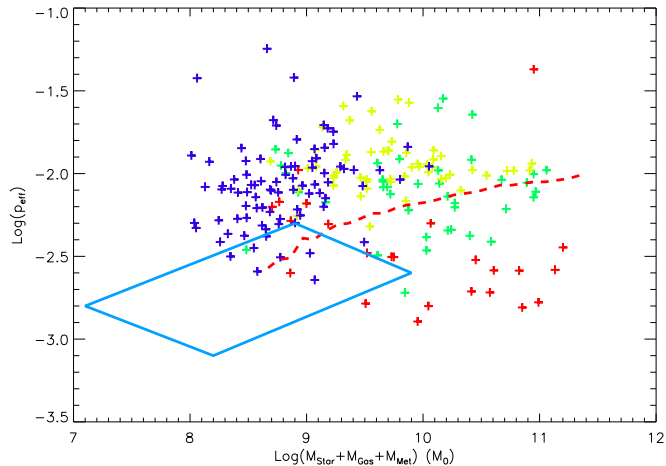}
\caption{Total baryonic mass against the effective yield ($p_{eff}$) - Red are galaxies with type earlier than Sa, green type Sa/Sb, yellow type Sc and blue galaxies later than Sc and dwarfs. The red dashed line is the median relation for $\sim53,000$ SDSS galaxies taken from Tremonti et al. (2004). The blue box represents the area occupied by the Virgo cluster dwarf irregular galaxies studied by Lee et al. (2003).} 
\end{figure} 

\begin{figure}
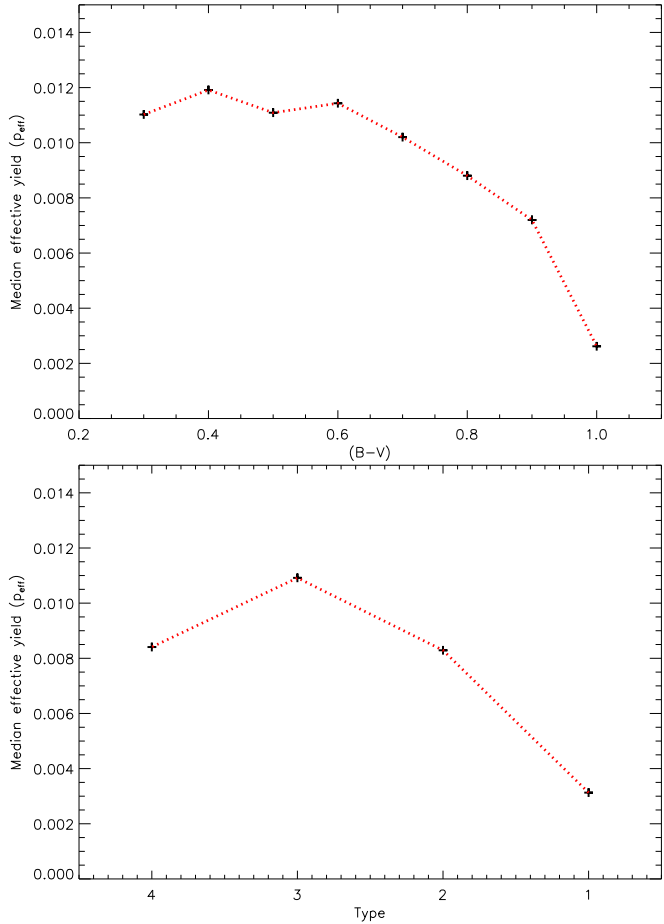

\centering
\includegraphics[scale=0.52]{colour_yield.epsi}
\includegraphics[scale=0.52]{type_yield.epsi}
\caption{Top - the effective yield against galaxy (B-V) colour. Bottom - the effective yield against galaxy type (1 is earlier than Sa, 2 Sa/Sb, 3 Sc and 4 dwarf/BCD/irr).} 
\end{figure} 

\subsection{The stellar mass star formation rate relation}
Using galaxies from the Herschel Reference Sample (HRS) Hughes et al. (2013) have applied the conversion relations of Iglesias-Paramo et al. (2006) ($\log{SFR_{NUV}} (M_{\odot} \mbox{yr}^{-1})=\log{L_{NUV}} (L_{\odot}) - 9.33$) to calculate the SFR of 39 of the 207 galaxies in our Herschel sample. Based on the assumption that the shortest far-infrared wavelength, and hence warmest dust, best correlates with the rate at which stars form (Calzetti et al. 2010), we have looked for a correlation between SFR and our 100$\mu$m data. In Fig. 18 we show that our 100$\mu$m luminosity correlates very well with the calculated SFR derived from the ultra-violet observations. To justify our use of the shortest wavelength data we have also looked at the correlation between the SFR and our longer wavelength far-infrared data (Fig. 18). The standard deviation of the data about the best fitting line in a log-log plot is 0.175, 0.183, 0.190, 0.190 and 0.200 for 100, 160, 250, 350 and 500$\mu$m respectively - the differences are not large, but the 100$\mu$m flux density does give the smallest scatter. The relation is $\log{SFR} (M_{\odot} yr^{-1})=0.73 \pm 0.05 \log{F_{100}} (Jy) - 17.1 \pm 1.1$.

We have then used this best fitting relationship to estimate SFRs for all 207 galaxies in our Herschel sample. Based on the assumption that our initial optical selection followed by far-infrared detection is not biased against star forming galaxies we will use this sample to define the SFR properties of Virgo cluster galaxies and of the cluster as a whole. 

In Fig. 19 (top) we have plotted the specific Star Formation Rate (sSFR) i.e. SFR per unit stellar mass, of our sample against the stellar mass. There is a clear trend for increasing sSFR with decreasing stellar mass. The trend is almost identical to that seen for field galaxies as indicated by the blue line on Fig. 19, which is the locus of the line derived by Schiminovich et al. (2007) using UV derived SFRs for $\sim 20,000$ galaxies (see their Fig. 7). In fact their line ($\log{\mbox{SFR}/M_{Star}}=-0.36\log{M_{Star}}-6.4$) is almost identical to that obtained by fitting our 39 SFR calibrating galaxies - each one indicated by a black box around each point on Fig. 19 ($\log{\mbox{SFR}/M_{Star}}=-0.30\pm0.07\log{M_{Star}}-6.8\pm0.7$). The red line on Fig. 19 (top) is also taken from Schiminovich et al. (2007, Fig. 7) and is the locus of what they describe as the non-star-forming sequence. Our conclusion is that our Virgo cluster galaxy sample generally fit the same sSFR stellar mass relation as is typical for galaxies in the local Universe. We have also distinguished the galaxies, as before, by their morphology. At higher stellar masses you move through S0 and earlier, Sa/Sb then to Sc as you move from lower to higher sSFRs - so sSFR depends not only on mass, but also morphology. The very late types/dwarfs have the highest sSFRs.

An alternative way of plotting the same SFR data is shown in Fig. 19 (bottom) - stellar mass against SFR. It is clear that the sSFR defines a timescale that is straight forwardly illustrated in the bottom plot in Fig. 19. The time scale is the time to form the current mass of stars at the current SFR - the star formation time scale. In Fig. 19 (bottom) a time scale of $10^{10}$ years is illustrated by the black dashed line. This line corresponds almost exactly with the locus of star forming galaxies (light blue line on Fig. 19) obtained using 100,000+ SDSS galaxies by Peng et al. (2010), their Fig. 1, and as can be seen is a reasonable fit to our Virgo cluster data. The two dotted lines are for timescales of $10^{11}$ and $10^{9}$ years for bottom and upper lines respectively. Galaxies probably undergo vast changes in their SFRs as they age, for example star bursts, but this plot does seem to distinguish galaxies of different types. The naive interpretation is one of younger age when going from red through green to yellow with the dwarf/irregular galaxies predominately young - consistent with what we said when discussing the chemical evolution model.

\begin{figure}
\centering
\includegraphics[scale=0.52]{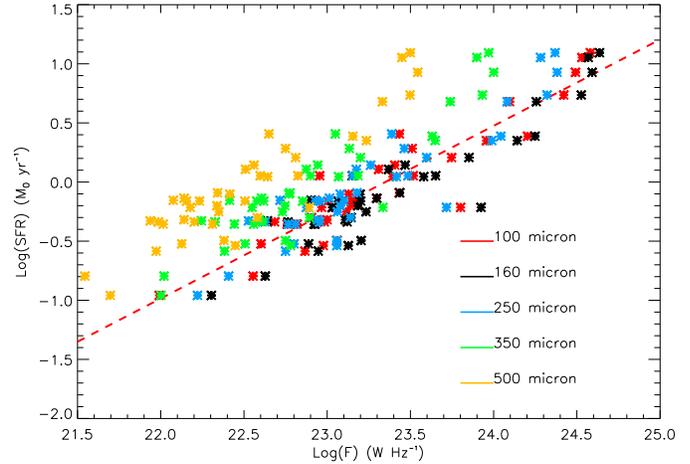}
\caption{The relationship between far-infrared luminosity in each band and their star formation rate. The red dashed line is the fit to the 100$\mu$m data, which has been used to infer SFRs for all 207 galaxies in the Herschel sample.} 
\end{figure}

\begin{figure}
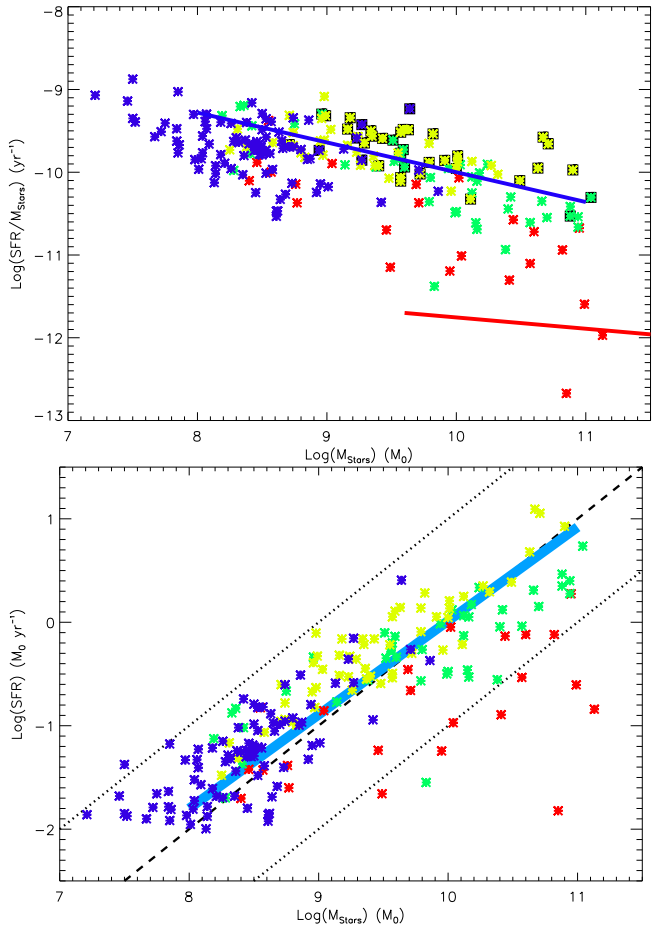

\centering
\includegraphics[scale=0.52]{sfr_mass_mass_2.epsi}
\includegraphics[scale=0.52]{sfr_mass.epsi}
\caption{Top - The SFR per unit mass of stars (sSFR) versus the mass of stars. The crosses mark the positions of the 207 galaxies in the Herschel sample. Galaxies are distinguished by their morphological type: Red are galaxies with type earlier than Sa, green type Sa/Sb, yellow type Sc and blue galaxies later than Sc and dwarfs. A black box around the cross marks those galaxies that were used to calibrate the 100$\mu$m SFR relation. The blue line marks the locus of the line of star forming galaxies and the red line that for passive galaxies (taken from Schiminovich et al. 2007, Fig.7). Bottom - The SFR versus the stellar mass for the 207 galaxies in the Herschel sample. The black dashed line indicates an age of 10$^{10}$ years with the lower and upper dotted lines a factor of 10 older and younger respectively. The thick blue line indicates the locus of star forming galaxies (taken from Peng et al. 2010, Fig. 1).} 
\end{figure}

Using the sum of the SFR of these galaxies and the cluster volume (62.4 Mpc$^{-3}$) used before, we estimate a cluster SFR density of 2.0 $M_{\odot}$ yr$^{-1}$ Mpc$^{-3}$. This compares with a local SFR density averaged over all environments of $\sim 0.03$ $M_{\odot}$ yr$^{-1}$ Mpc$^{-3}$ (Robotham and Driver 2011). The cluster is an over density in SFR by a factor of $\sim66$ compared to the local mean value. This is almost a factor of two lower than the stellar mass over density of $\sim 110$ (Table 3) - a reflection of the increased numbers of quiescent galaxies in the cluster environment. 

Multiplying the derived SFR density by the characteristic age of $\sim10^{10}$ yr gives a stellar mass density of $2.0 \times 10^{10}$ M$_{\odot}$ Mpc$^{-3}$ compared to that calculated from the integral of the luminosity function of $3.3 \times 10^{10}$ M$_{\odot}$ Mpc$^{-3}$ (Table 5). So the SFR must have been on average marginally, but not considerably, higher in the past. Given the value of $\lambda$ calculated in sub-section 8.1 the chemical evolution model predicts a gas mass loss rate density of $1.4 \times 2.0=2.8$ M$_{\odot}$ yr$^{-1}$ Mpc$^{-3}$, which amounts to $2.8 \times 10^{10}$ M$_{\odot}$ Mpc$^{-3}$ of material deposited in the intra-cluster medium over 10$^{10}$ years. 

Dividing the SFR density by the mass density of stars gives a sSFR for the cluster as a whole of $6.1 \times 10^{-11}$ (-10.2) yr$^{-1}$ (where the value in brackets is the $\log{}$ for comparison with Fig. 19). Using the Robotham and Driver (2011) value for the local SFR density and the Panter et al. (2007) value for the local mass density of stars we get a sSFR for the field of $1.0 \times 10^{-10}$ (-10.0) yr$^{-1}$. As already demonstrated in Fig. 19 this comparison of the current sSFR of the Virgo cluster and the local field gives no indication of any dramatic difference between the field and cluster that might be due to environmental effects.

\begin{figure}
\centering
\includegraphics[scale=0.52]{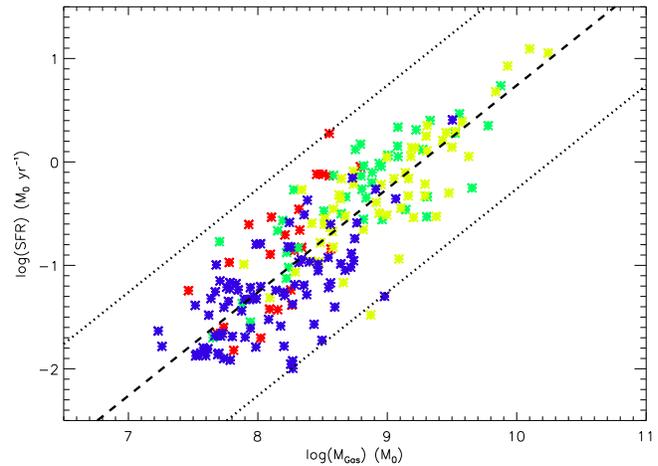}
\caption{The SFR versus the gas mass for the 207 galaxies in the Herschel sample. The black dashed line indicates a gas depletion time scale of $1.8 \times 10^{9}$ years with the upper and lower dotted lines a factor of 10 shorter and longer respectively. Galaxies are distinguished by their morphological type: Red are galaxies with type earlier than Sa, green type Sa/Sb, yellow type Sc and blue galaxies later than Sc and dwarfs.} 
\end{figure}

We can also plot the SFR against the gas mass (in this case HI + H$_{2}$) of our Herschel sample galaxies, Fig. 20. This is effectively the 'global' Schmidt/Kennicutt law (Schmidt, 1959, Kennicutt, 1998), which usually relates the gas surface density ($\Sigma_{Gas}$) to the star formation rate surface density ($\Sigma_{SFR}$). Typically and locally within galaxies it is found that $\Sigma_{SFR}=A\Sigma_{Gas}^{N}$ where $N \approx 1.4$ (Kennicutt, 1998). Our 'global' relation, shown in Fig. 20, has a flatter slope than this with an almost linear relation between gas mass and SFR ($N_{Global}=0.96 \pm 0.04$). This 'global' relation may actually be of more significance than that between SFR and the mass of stars shown in Fig. 19 (Bigiel et al. 2008). There, because of the large scatter in the data, we just illustrated a line of age $10^{10}$ years, but Fig. 20 shows a much better correlation between the data. With a linear relation between gas mass and SFR the gas specific star formation rate ($SFR/M_{Gas}$) is approximately constant for all galaxies and can be expressed as $\sim 6 \times 10^{-10}$ M$_{\odot}$ converted into stars each year for each solar mass of gas. The fit to the data also defines the gas depletion time scale (time to consume the gas at the present SFR) which at $10^{9}$ years is a factor of 10 shorter than the star formation time scale defined above. The star formation time scale describes where a galaxy has been, while the gas depletion time scale defines where it is going. So, a quantity of interest is the ratio of the gas depletion to the star formation time scale for galaxies of different morphological types. The median ratio is 0.04, 0.08, 0.38 and 0.51 for our four morphological types, earlier than Sa, Sa/Sb, Sc and later than Sc respectively. This quantifies a morphological age sequence with those types earlier than Sa at the end of their star forming lives, while those types later than Sc in their middle age.

\subsection{The stellar mass metallicity relation}
In section 7 we used the mass metallicity relation to predict metalicities for galaxies in our sample that did not have a measured oxygen abundance. Here we want to briefly discuss the relationship itself. We will not dwell on this point because to some large extent this has already been discussed by us in Hughes et al. (2013). Briefly, there are two major issues with regard to the stellar mass metallicity relation - its origin and whether it is different in different environments. With regard to its origin the most common scenario is gas loss due to stellar winds in galaxies of low mass, while larger galaxies retain their gas. This is difficult to sustain within the bounds of our sample as we have already shown that we do not find a global relationship between effective stellar yield ($p_{eff}$) and baryonic mass - on average our low mass galaxies do not have lower values of $p_{eff}$ commensurate with gas loss (Fig. 16). As before we suggest that this might be a selection effect, as we have selected galaxies that have as yet not been subject to gas loss. 

Having a sample that does not have a lower value of $p_{eff}$ for low mass galaxies yet still has a mass metallicity relation leads one to suggest that mass loss is not the origin of the mass metallicity relation. Instead we suggest that what we are seeing is a sequence of age. Low mass galaxies have large gas fractions (Fig. 15), high specific star formation rates (Fig. 19, top) and young ages (Fig 19, bottom). 

With regard to the second issue and in agreement with our conclusions in Hughes et al. (2013) we find little evidence for a difference in the stellar mass metallicity relation with environment. Our data is shown in Fig. 13 (bottom) with a linear least squares fit indicated by the dashed black line. Note our sample is larger than that used by Hughes et al. (2013) because we have Virgo galaxies observed by Hughes et al. (2013), that were not in their primary sample (HRS). The red dashed line is the mean relation for non-cluster galaxies taken from Hughes et al. (2013). Although this line apparently sits below our relationship for cluster galaxies, Hughes et al. (2013) use a different prescription for calculating stellar mass (Salpeter rather than Kroupa IMF), which we estimate shifts the dashed red line to the left by $\approx$0.3 dex. Given the scatter in metallicity about the line of $\approx0.1$ dex this makes the two lines consistent with each other. We conclude, as in Hughes et al. (2013), that there is no evidence for a higher metallicity at a given stellar mass of cluster compared to field galaxies.

\subsection{The baryonic Tully-Fisher relation}

\begin{figure}
\centering
\includegraphics[scale=0.52]{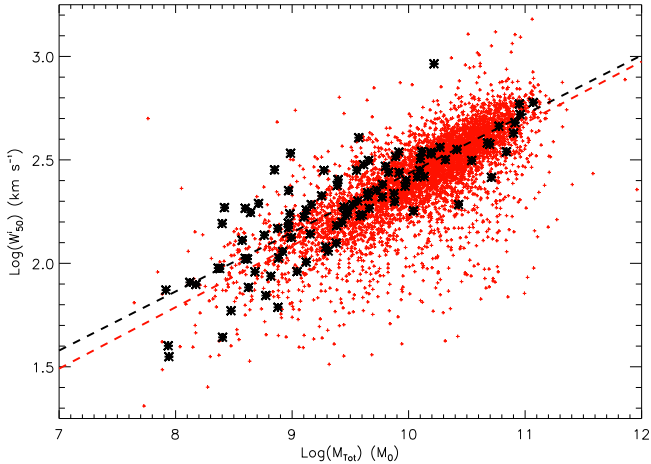}
\caption{The baryonic Tully-Fisher relation. The black crosses are for 100 Virgo cluster galaxies from the Herschel sample. The red dots are for 5174 galaxies taken from the LEDA database. $M_{Tot}$ is the total baryonic mass for the Virgo galaxies and the stellar plus gas mass for the Leda galaxies. $W^{i}_{50}$ is the inclination corrected velocity width at 50\% of the peak flux density for the Virgo galaxies and twice the maximum velocity for the LEDA galaxies. } 
\end{figure} 

One of the most studied, well defined and used scaling relations for galaxies is the Tully-Fisher relation. The relation was originally used as a means of obtaining distances to galaxies independently of their redshift in order to measure peculiar velocities. It has since been extensively used by numerical simulators of galaxies and large scale structure to equate something that they measure in their simulation (rotation) to what is observed (luminosity). What is still elusive is the precise physical origin of the Tully-Fisher relation. 
Given that the Tully-Fisher relation is between velocity and luminosity the simplest assumption is that luminosity is acting as a proxy for mass (or possibly some combination of mass and size). McGaugh et al. (2000) showed that by using the total baryonic mass (stars plus gas) of a galaxy in place of the luminosity the scatter in the relation is much reduced - this is known as the baryonic Tully-Fisher relation. 

For 100 galaxies from the 207 in the Herschel sample we can obtain from ALFALFA a 21cm line width (width at 50\% of peak flux density - $W_{50}$) and from NED semi-major ($a$) and semi-minor ($b$) axes sizes (measured at the 25th blue magnitude isophote). The axis ratio can be used to obtain the inclination ($\sin{i}=\sqrt{\frac{1-(b/a)^{2}}{1-0.15^{2}}}$, Stark et al. 2009) and so correct the measured to the intrinsic velocity width ($W_{50}^{i}$). The total baryonic mass ($M_{Tot}$) is just the sum of the mass in stars, gas and metals we used earlier.  The relation we obtain is shown in Fig. 21 (black crosses). The gradient of the line (black dashed line) is measured to be $0.29\pm0.02$. This value is reasonably consistent with previously derived values of 0.25 by Stark et al. (2009) and 0.31-0.33 (sample dependent) by Gurovich et al. (2010).

To make a comparison to a data set that samples galaxies over a wide range of environments, not just a cluster, we have selected all 5174 galaxies (Sab or later) from the Lyon extra-galactic database (LEDA) that have an I band magnitude, a gas mass (atomic hydrogen) and a 'maximum' rotation velocity ($v_{m}$). For each LEDA galaxy we have then simply obtained a stellar mass using an absolute I band magnitude for the Sun of $M^{I}_{\odot}=4.08$, a total mass ($M_{Tot}$) by summing the stellar and gas mass and equating $W_{50}^{i}$ to $2 \times v_{m}$. The data for the LEDA galaxies is shown on Fig. 21 as red dots. 

Comparing the cluster and non-cluster data there appears to be no evidence that the baryonic Tully-Fisher relation is any different for Virgo cluster and non-cluster galaxies. The slope of the line fitted to the LEDA data is $0.297\pm0.004$ (red dashed line) consistent with that for the cluster galaxies and what has been derived before.

\subsection{The mass size relation}

\begin{figure}
\centering
\includegraphics[scale=0.52]{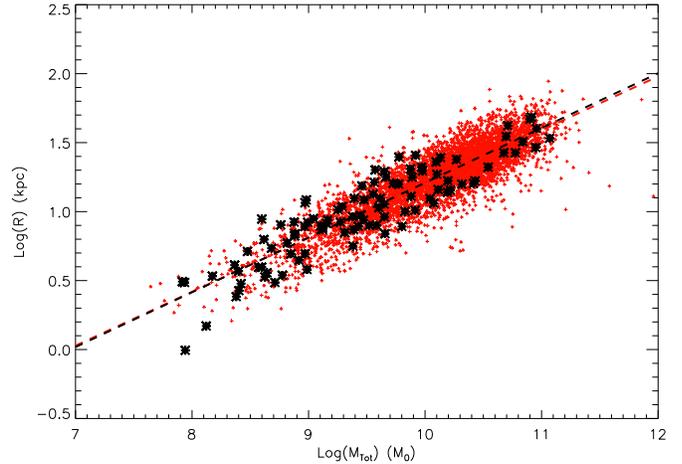}
\caption{The baryonic mass size relation. The black crosses are for 100 Virgo galaxies from the Herschel sample. The red dots are for 5174 galaxies taken from the LEDA database. $M_{Tot}$ is the total baryonic mass for the Virgo galaxies and the stellar plus gas mass for the Leda galaxies. $R$ is the radius measured along the major axis at the 25th B$\mu$ isophote.} 
\end{figure} 

Another often quoted scaling relation of galaxies is that between mass and size. Given the data we have for the 100 Virgo cluster galaxies described above we can also plot the radius (semi-major axis) against the total baryonic mass ($M_{Tot}$), Fig. 22. The Virgo galaxies again show a good correlation between these two quantities with a measured slope of $0.40\pm0.02$. This compares with a value of 0.32 obtained by Avila-Reese et al. (2008) for a smaller sample of field galaxies.

Again we can use the LEDA data as a comparison sample - red dots Fig. 22. The measured slope of the relation for the LEDA galaxies is $0.390\pm0.003$ consistent with the Virgo cluster sample. There is no evidence that the cluster environment has altered in any way the global relationship between mass and size - something that might have been expected if, for example, the tenuous outer regions of cluster galaxies were to fall prey to gravitational interactions. 

\subsection{Comments on the implications of the scaling relations}
A typically subscribed to explanation of the scaling of the baryonic mass-velocity (T-F) relation (Gurovich et al. 2010, and references therein) starts with an isothermal sphere for which the density is: $\rho(r)=\frac{v^{2}}{4 \pi G r^{2}}$. Then motivated by numerical simulations, regions that collapse into a virialised halo have conventionally a size of $r_{200}$, which is the size that contains a mean density 200 times that of the critical density. Substituting in this density and size and using $M=\frac{4}{3}\pi\rho_{200}r^{3}_{200}$ leads to $v \propto M^{1/3}$ - very close to what is observed. 

However, there are a number of caveats we should add to this. Firstly, it is not clear how this explanation fits in with the currently favoured galaxy formation model, which relies primarily on the growth of structure through merging of smaller masses and not the collapse of an individual isothermal sphere. Secondly, the argument relies on galaxies all forming from regions of the same initial density, variations in the initial density will produce scatter in the plot and possibly a power law index not quite equal to 1/3. Thirdly, there are other means (see below) of obtaining this power law relationship. 

With regard to the mass-size relation the above reasoning leads to $r \propto M^{1/3}$ directly - very close to what is observed - but changes in the initial density will again lead to scatter in the relation. In addition we also have to accept that a spherical halo of size $r_{200}$ scales linearly to a disc of size $r_{D}$ and that this all happens while each galaxy undergoes substantial merger activity. There are also other ways of obtaining this power law relationship.

The mass metallicity and the mass star formation rate relations also demonstrate the fundamental role that galaxy mass plays in both the chemical evolution and star formation history of galaxies. Given these four relations, that depend so critically on the mass of a galaxy, one is tempted to surmise that once a galaxy's mass has been laid down almost everything else follows in a well defined way. As the scaling relations described above seem to apply equally well to cluster and field galaxies it is clear that internal evolutionary processes dominate over those inflicted by the local environment.

The above is in many ways a restatement of the conclusions of Disney et al. (2008) who showed that out of six primary properties of galaxies five correlated very strongly with one other (mass?). They stated that: "Such a degree of organization appears to be at odds with hierarchical galaxy formation, a central tenet of the cold dark matter model in cosmology.".  

An alternate scenario for individual galaxies, though maybe not such a good explanation of large scale structure, is that galaxies actually do form from the collapse of approximately isothermal spheres. The collapse of an isothermal sphere, in such a way as to preserve its angular momentum distribution as a function of mass, leads to a disc with a $1/r$ surface density distribution and a flat rotation curve (A Mestel disc - Mestel, 1963). As long as the initial isothermal spheres have the same density then scaling parameters of such a disc depend only on the initial mass such as $r \propto M^{1/3}$ and $v \propto M^{1/3}$. These relations are the same as those predicted above and are very close to what is observed, but in addition they are much more precisely defined. For example, the Mestel disc parameters ($M$, $r$ and $v$) are the same throughout the collapse of the spherical cloud into a disc and there is no issue with mergers (Davies, 2012). In addition there seems to be an age sequence with more massive early types older than the less massive later types - it is difficult to see how this can be accommodated within the current hierarchical picture of galaxy formation.

\section{Material in the intra-cluster medium}
As we will see, considering just the galaxies detected in multi-wavelength surveys is really just scratching at the surface of the cluster baryon content. In this section we will review, assess and use data from surveys that have attempted to measure the mass of baryonic material between the galaxies.

Given our expectations of gravitational interactions between the galaxies and with the overall cluster potential we might expect to find galaxy debris in inter-galactic space. It is also not clear how efficient the galaxy formation process is. For example is star formation confined to the bright easily identifiable galaxies or is there a population of faint stellar systems between the prominent concentrations of stars? And how efficiently has the total cluster gas been taken into the individual stellar systems? 

 Searches for an intra-cluster stellar population have previously been made by either looking for the diffuse low surface brightness signal of inter-galactic light or by trying to identify individual stars. Beautiful images of the diffuse light in Virgo have been made by Mihos et al. (2005). It seems that this diffuse light is concentrated in extended haloes around the cluster galaxies with the addition of some filamentary structures between them. The first direct detection of individual stars between the galaxies was made by Smith (1981) when he observed a type Ia supernova in the region between M86 and M84. By comparing source number counts Ferguson et al. (1998) have inferred the presence of red giant stars in the inter- galactic medium. Several groups have used narrow band imaging to detect inter-galactic planetary nebulae (Arnaboldi et al., 2002, 2003), Feldmeier et al., 1998, 2003). In Mihos et al. (2009) the inter-galactic light and planetary nebulae methods are compared to see whether both are tracing the same stellar structures in the inter-galactic medium - the result being '..a rough correspondence on large scales ($\sim 100$ kpc)'. Arnaboldi (2003) concludes that about 20\% of the light in individual cluster galaxies is produced by inter-galactic stars. By simply converting this directly to a mass ($M/L=1$) we have a mass density of stars outside of galaxies of $\sim 7 \times 10^{9}$ $M_{\odot}$ Mpc$^{-3}$ a fraction of what is in the galaxies ($3.3 \times 10^{10}$ M$_{\odot}$ Mpc$^{-3}$, table 5). Hopefully new surveys like the Next Generation Virgo Cluster Survey (NGVCS, Ferrarese et al., 2012) will reveal more about the nature of the stars between the bright galaxies.

Even more so than stars we might expect gas to be removed from galaxies as they move through the cluster environment, for gas is not only affected by gravity, but also by stellar winds and ram pressure stripping by the X-ray gas. Blind surveys for atomic hydrogen in the Virgo cluster have found very little when compared to what is in galaxies. Davies et al. (2004) carried out a blind survey over 32 sq deg of the cluster and found just 2\% of the HI detected was in previously unidentified sources, predominantly in the form of tidal streams. A similar conclusion has been made using the larger area survey of Kent et al. (2007) (see also Kent et al. 2009). We know of no survey that has put limits on the molecular hydrogen and helium mass of material in the inter-galactic medium and so adjust for these components in the same way as we have before ($1.7M_{HI}$). This leads to a mass density of cold/cool gas outside of galaxies of $\sim 9 \times 10^{7}$ $M_{\odot}$ Mpc$^{-3}$. Again a small fraction of what is in the galaxies ($4.3 \times 10^{9}$ M$_{\odot}$ Mpc$^{-3}$, Table 5).

Putting the above rough estimates together the ratio of mass of stars to mass of atomic gas is about 10 times higher in the inter-galactic space than in the prominent galaxies. Previously we have argued that there is good evidence for gas stripping of cluster galaxies and little evidence for the removal of stars i.e. gas is depleted compared to stars and dust. So, our simple expectation would be just the opposite to what is observed - relatively more gas than  stars in the inter-galactic medium. Either the inter-galactic stars have formed in situ (from stripped or primordial gas) or at sometime in the past gravitational interactions were much more efficient. 

However, the relatively cold atomic gas drawn out via stripping processes is but a small fraction of the mass of gas that resides between the galaxies. By far the largest contribution to the baryon density in both the inter-galactic medium and the cluster as a whole is that due to the hot (X-ray) gas (Bohringer et al. 1994, Urban et al. 2011). It is generally assumed that this gas has been expelled from galaxies by stellar (super novae) winds (Bohringer 2004), hence its high temperature. Observations using ROSAT have shown that most of the hot gas is concentrated around the three bright galaxies M87, M86 and M49 with about 83\% of the gas mass in the extended halo around M87 and only about 15\% in a more diffuse 'cluster' component. The total mass density of hot cluster gas is calculated to be $\sim 10^{12}$ $M_{\odot}$ Mpc$^{-3}$ (we have used 20\% of the gravitational mass measured by the X-rays, see Bohringer et al. 1994 for further details). This hot X-ray gas completely dominates the baryon budget with about a factor of twenty-five times more mass in this component than in the stars and cool gas in galaxies and that outside of galaxies combined.

We know of no observations that conclusively show that dust resides in the inter-galactic medium of the Virgo cluster
\footnote{Stickel et al. (1998) have previously presented evidence for inter-galactic dust emission from the Coma cluster.}, though Cortese et al. (2010) conclude that dust is being stripped as well as gas from the Virgo cluster galaxy NGC4438. We will be using the HeViCs, along with other data (21cm and the reddening of background galaxies), in the future to try and address this issue. In any case given the dominance of the hot gas we do not expect the metals in inter-galactic dust to amount to very much (comparatively). Bohringer et al. (1994) give a metallicity value for the hot gas as $\sim 0.5z_{\odot}$. This leads to a mass density in metals of $\sim 7 \times 10^{9}$ $M_{\odot}$ Mpc$^{-3}$, which is about 200 times more than the mass of metals in the galaxies. This is all summarised in Table 6.

\begin{table}
\begin{center}
\begin{tabular}{c|cccccc}
Baryonic              & $\rho$ \\
Component             & ($M_{\odot}$ Mpc$^{-3}$) \\ \hline
 Total cluster stars  & $4\times10^{10}$ \\
 Total cluster gas    & $1\times10^{12}$ \\
 Total cluster metals & $7\times10^{9}$  \\
\end{tabular}
\caption{The total cluster mass within its various components.}
\end{center}
\end{table}

Our view of the chemical evolution of the cluster as a whole is now completely changed. In the previous section we concluded that to be consistent with a simple closed box chemical evolution model, the cluster galaxies must have lost as gas about 1.4 times their stellar mass to the inter-galactic medium. In fact when we do the accounting we find about a factor of twenty-five times more gaseous material in the intra-cluster medium than in the stars. This is shown by the position of the left large black star in Fig. 15 (using the total cluster gas fraction and metallicity from Table 6). The closed box model interpretation of this is that there must have been substantial inflow of enriched gas to get this metallicity at such a high gas fraction. Of course the closed box chemical evolution model is not really applicable to the cluster as a whole because the hot gas is not available for continued star formation. 

So, how can we interpret the position of the left large black star on Fig. 15 given the White et al. (1993) assertion that baryons are retained within the cluster? If we accept that the galaxies have lost about 1.4 times their stellar mass as out flowing gas then only about 5\% of the intra-cluster X-ray gas can have its origin from within galaxies and the remaining 95\% must have a "primordial" origin. This is consistent with the numerical models of Matteucci and Gibson (1995) who say that only 1-10\% of the intra-cluster gas can have originated from the galaxies. The immediate conclusion is that within a cluster like Virgo the galaxy formation process has been very inefficient with only about 5\% of the baryonic mass ending up in galaxies. How about the total metallicity of the cluster - can the total mass in metals have been produced by the stars contained within the galaxies? Given the intra-cluster mass density of metals ($7.0 \times 10^{9}$ M$_{\odot}$ Mpc$^{-3}$, for a X-ray gas metallicity of 0.5$z_{\odot}$, Bohringer et al., 1994) and the predicted mass of out flowing gas ($1.4 \times 3.3 \times 10^{10}$ M$_{\odot}$ Mpc$^{-3}$, Table 5) we require a gas outflow from the galaxies with a super solar metallicity of about 13$z_{\odot}$. Super solar metalicities like this are possible if the wind consists primarily of supernovae ejecta. 

To assess if this is a viable explanation of the intra-cluster metals we consider a simple model of metal enrichment due to galactic supernovae winds. We have used a power law stellar initial mass function (IMF) between 0.1 and 100 M$_{\odot}$ (Matteucci and Gibson, 1995) and the type II supernovae yields given by Arnett (1991) to predict the mass density of metals produced by supernovae driven galactic winds resulting from the observed mass density of stars. We derive the (O/H) ratio and then calculate the total mass in metals assuming solar abundances. For a Salpeter (1955) IMF power law slope of -2.35 we predict a metals mass density of $3.9 \times 10^{8}$ M$_{\odot}$ Mpc$^{-3}$ and for a less steep Arimoto and Yoshi (1987) IMF power law slope of -1.95, $1.4 \times 10^{9}$ M$_{\odot}$ Mpc$^{-3}$ - this is for 100\% efficient supernovae mass loss. Neither of these IMFs can account for the metals found in the intra-cluster medium and even with a IMF slope of zero there are insufficient supernovae produced metals to account for the observed metals mass density of $7.0 \times 10^{9}$ M$_{\odot}$ Mpc$^{-3}$. We conclude that the 'primordial' intra-cluster X-ray gas must also have been enriched to some extent by stars other than those found in the galaxies. 

We note that the above assumes that the enrichment of the intra-cluster medium only occurs due to type II supernovae. This is consistent with the findings of Matteucci and Gibson (1995) who show that later galactic winds due to evolved low mass stars and type Ia supernovae generally have insufficient energy to deposit material in the intra-cluster medium (at most 30\% of that deposited by type II supernova). If 
$1.4 \times 10^{9}$ M$_{\odot}$ Mpc$^{-3}$ is taken as an upper limit on the mass of metals expelled by galaxies then a lower limit on the metallicity of the 'primordial' gas is $\sim0.4z_{\odot}$. This is far higher than is expected from models of population III star gas enrichment which are predicted to be $\sim10^{-4}z_{\odot}$ (Kulkarni et al. 2013, Trenti et al. 2009) and the metallicity of population II stars, which formed from the population III enriched gas. 

We are thus left with two interesting observations: the gross inefficiency of the galaxy formation process (only 5\% of the available baryons end up in galaxies) and the relatively large abundance of metals in this discarded gas. Our result is not controversial as our derived gas to total baryons mass fraction of $\approx 0.04$ is consistent with models of cluster formation (Planelles et al. 2013) and with observations of other clusters (Vikhlinin et al., 2006). What is difficult to understand is the metallicity of the gas.

If galaxy formation is this inefficient everywhere then the stellar mass density needs to be multiplied by a factor of $\sim$25 to get the total baryonic mass. Using the field stellar mass density given in Table 3 of $3.0 \times 10^{8}$ M$_{\odot}$ Mpc$^{-3}$ leads to a predicted baryon density of $8.0 \times 10^{9}$ M$_{\odot}$ Mpc$^{-3}$ due to this inefficiency. This compares consistently with a baryon density of $6.0 \times 10^{9}$ M$_{\odot}$ Mpc$^{-3}$ derived from the cosmological model (Komatsu et al., 2011). If this universal inefficiency is true then most of the baryons involved in the galaxy formation process remain hidden - more evidence for a warm inter-galactic medium (Cen and Ostriker, 1999) - except in clusters where the large gravitational potential reveals them through their X-ray emission. The origin of the intra-cluster metals is not clear - they cannot have been synthesised by the stars in galaxies or by a generation of population III stars. If the origin of the intra-cluster gas is connected with AGN rather than star formation activity then the above conclusions will need to be revisited.

\begin{figure*}
\centering
\includegraphics[scale=0.9]{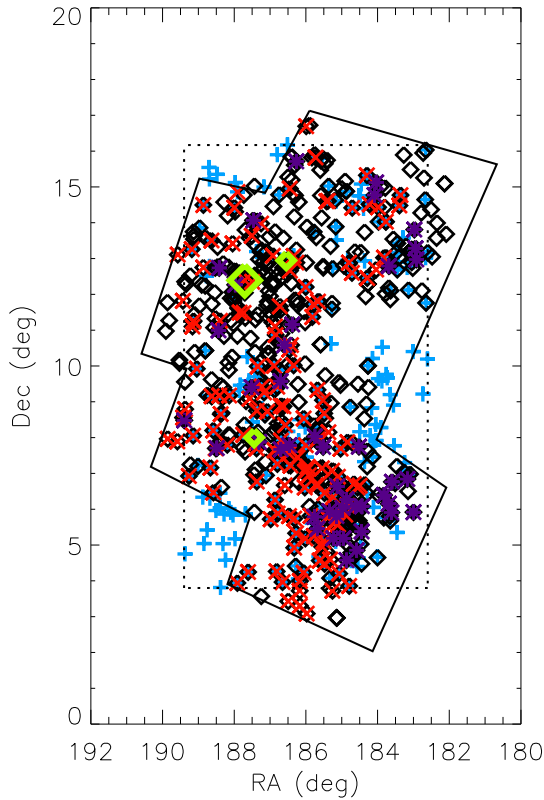}
\caption{The positions on the sky of the galaxies used to measure the dynamical mass. Black diamonds - the VVC+ sample, blue crosses the HI sample, red crosses the Herschel sample, purple diamond the new SDSS optical detections. The large green diamond marks the position of M87 and the upper and lower smaller green diamonds the positions of M86 and M49 respectively. The solid black line shows the approximate outline of the region of sky observed by Herschel. The black dotted line the region extracted from the HI ALFALFA data.} 
\end{figure*} 

\section{Dark matter and cosmological implications}
We can obtain an estimate of the total cluster mass in two ways. Firstly, by using the X-ray data to measure both the hot gas mass density and temperature profiles and then assume hydrostatic equilibrium (Briel et al. 1992). Secondly, by using the line-of-sight velocities of the galaxies and assuming virial equilibrium. Neither of these assumptions, hydrostatic or virial equilibrium, are necessarily true. Clusters including Virgo are observed to have cooling flows within their inner regions and although the hot gas mass distribution is smoother than that of the galaxies it is still clumped on galactic scales (Bohringer et al. 1994). As described in the introduction dynamically Virgo, along with other clusters, seems to consist of a number of galaxy sub-clusters, clouds, groups and individual galaxies that are currently falling into the cluster and hence have not settled into a virialised mass distribution. Hence they are not good test particles for measuring the cluster gravitational potential (Binggeli et al. 1987, Binggelli et al. 1993) - note also the two 'quantised' distances of 17 and 23 Mpc we have used above for cluster members. However, we will use both the X-ray and dynamical data to try and provide some insight into the total cluster mass.

Bohringer et al. (1994) show that the X-ray halo of M87 can be traced out to about 1.5 Mpc ($D_{Virgo}$=17 Mpc) or 5 deg and that the majority of the gravitating mass, as measured by the X-ray gas, is within this distance. It is clear that the majority of the X-ray emission actually comes from within about 2 deg of M87 (see Fig. 1 in Bohringer et al. 1994) well within the extent of our Herschel data. Based on the assumption of hydrostatic eqillibrium Bohringer et al. (1994) calculate a gravitational mass for the halo around M87 of $1.5-6.0 \times 10^{14}$ M$_{\odot}$ - we will take this to be the mean value of $3.8 \times 10^{14}$ M$_{\odot}$. Much smaller halos around M86 and M49 are calculated to have masses of about $2.0 \times 10^{13}$ M$_{\odot}$. This gives a total gravitating mass, as measured by the hot X-ray gas, of $\sim 4.2 \times 10^{14}$ M$_{\odot}$. 

 Given the apparent dominance of M87 in the X-ray data we can regard the cluster galaxies as test particles moving in its gravitational field. In this case the cluster mass is given by $M_{cluster}=\frac{3\pi}{2G}R_{e}\sigma_{v}^{2}$ where $R_{e}=\frac{N}{\sum_{i=1}^{N}(|R_{i}-R_{M87}|)^{-1}}$ is the effective radius of the cluster and $\sigma_{v}^{2}=\frac{\sum_{i=1}^{N}(v_{i}-v_{M87})^2}{N}$ is the velocity dispersion of the galaxies (Binney and Tremaine, 1994).   We can use the ALFALFA data we used earlier (261 galaxies with $400<v_{helio}<2659$ km s$^{-1}$) to make an estimate of the dynamical mass of the cluster. Using these galaxies we derive a value of $R_{e}=1.8$ Mpc and with a measured $\sigma_{v}$ of 616 km s$^{-1}$ a cluster mass of $M_{cluster}=4.7\times10^{14}$ M$_{\odot}$, consistent with the X-ray data. Our mean cluster velocity is measured to be 1552 km $s^{-1}$ considerably higher than that given in NED for the cluster (1079 km s$^{-1}$) and for the measured velocity of M87, which is 1258 km $s^{-1}$ (Binggeli et al. 1993). Our exclusion of galaxies with low ($<400$ km s$^{-1}$) and negative velocities biases our value compared to other derivations that include them (Binggeli et al. 1993). However, it is the velocity dispersion and not the mean velocity that is important here. Previous measurements of the cluster velocity dispersion give a value of $\sigma=721$ km s$^{-1}$ (Ferguson and Sandage, 1990), a little higher than our value of 616 km s$^{-1}$. The cluster mass is given in Ferguson and Sandage (1990) as $M_{cluster}=5.2\times10^{14}$ M$_{\odot}$ when adjusted to a distance of 17 Mpc and as $M_{cluster}=4.2\times10^{14}$ M$_{\odot}$ when modelled as a NFW halo (McLaughlin, 1999). All these values along with that measured by the X-rays are consistent with the value of $M_{cluster}=4.7\times10^{14}$ M$_{\odot}$ we measure here using the ALFALFA sample galaxies.

As an aside we can also use the stellar and dust samples to measure the dynamical properties of the cluster - 546 galaxies with optical (SDSS) velocities and 146 galaxies in the Herschel sample that have HI velocities from ALFALFA. The results are summarised in Table 7 and the distributions of velocities shown in Fig 24. All three derived cluster masses are consistent with those given above. Given the different spatial distributions of gas rich and gas poor galaxies (morphology density relation) it is surprising that they have almost exactly the same dynamical characteristics, particularly as it is thought that the cluster is being assembled from its in falling gas rich galaxy population (Boselli and Gavazzi, 2006).

The derived effective size of the cluster ($R_{e}$) is approximately 2 Mpc (Table 7). Assuming a spherical distribution this gives a cluster volume of 33.5 Mpc$^{-3}$ compared to the value of 62.4 Mpc$^{-3}$ used above. Put another way we have effectively used a value of $\sim1.2R_{e}$ when calculating the cluster volume.
 
Galaxies with $400<v_{Helio}<2600$ km s$^{-1}$ are used to construct the 'stars' and 'gas' samples shown in Fig. 24 and what is obvious is the bi-modal structure obtained by including galaxies from both the sub-clusters and clouds - as described in the introduction. What is not so obvious is why this bi-modality should also still be present in the Herschel data which is restricted to galaxies classified in GOLDMINE as in sub-clusters A and B - either there are some interlopers or the tail of the distribution of sub-cluster velocities extends to velocities that overlap with the clouds. 

That the X-rays and the galaxies measure the same gravitational potential can also be checked by comparing the X-ray temperature with that predicted from the velocity dispersion of the galaxies i.e. $T=\frac{m_{p}\sigma_{v}^{2}}{k_{B}}$ (Felton et al., 1966). This gives a value of $T=3.7 \times 10^{7}$ K, reasonably close to the measured X-ray temperature of $\sim T=2.6 \pm 0.3 \times 10^{7}$ K (Urban et al. 2011).
 
With a size of $R_{e}=1.8$ Mpc and a velocity dispersion of $\sigma_{v}=616$ km s$^{-1}$ the cluster crossing time is of order $3 \times 10^{9}$ years. This is about 1/40th of the Hubble time, 1/3rd of the star formation time scale, 3 times the gas depletion time and about 10 times the typical rotation period of a galaxy. 

The distribution on the sky of the galaxies used to measure the dynamical mass is shown in Fig. 24. Black diamonds are the 546 galaxies with velocities taken from the VCC+ sample, blue crosses are the 261 galaxies taken from the HI ALFALFA sample, red crosses are the 146 galaxies with velocities taken from the Herschel sample. The large green diamond marks the position of M87, which has been used as the dynamical centre of the cluster. The upper and lower smaller green diamonds mark the positions of M86 and M49 respectively - M87, M86 and M49 are all centres of X-ray emission.

Given a total cluster baryon mass density of about $1.0 \times 10^{12}$ M$_{\odot}$ Mpc$^{-3}$ the cluster is about a factor of $\sim150$ more dense in baryons than the cosmologically measured value of $\Omega_{b}=0.046$ ($\rho_{Critical}=1.4 \times 10^{11}$ M$_{\odot}$ Mpc$^{-3}$). Given the total dynamical mass density of about $8 \times 10^{12}$ M$_{\odot}$ Mpc$^{-3}$ the cluster is about a factor of 200 more dense in all forms of matter than the cosmologically measured value of $\Omega_{m}=0.272$ (the above cosmological values are taken from Komatsu et al., 2011).
So, the ratio of dynamical (total) to baryonic mass ($f_{d}$) for the cluster as a whole is about 8 compared to a value of about 6 for the ratio of total to baryonic matter in the Universe as a whole. Although probably consistent within the impossible to estimate errors higher values of $f_{d}$ for clusters when compared to the cosmic value have previously been reported by McCarthy et al. (2007). They obtain a mean value for galaxy clusters of $f_{d}=8$ the same as we derive here.

We can use our cluster galaxy data to measure this dynamical to baryonic mass ratio for individual galaxies. As stated earlier, of the 207 galaxies in our Herschel selected sample for which we have measured total baryonic masses there are 100 which have both a measured size ($R_{ISO}$), inclination ($i$) and a HI velocity width ($W_{50}$). To estimate individual dynamical masses we use: $M_{dyn}=\frac{R_{ISO}W^{{i}^{2}}_{50}}{{G}}$, where $R_{ISO}=\sqrt{r_{minor}r_{major}}$ and $W_{50}^{i}$ is the inclination corrected velocity width (Cortese et al. 2008). The distribution of dynamical over baryonic mass for these 100 galaxies is shown in Fig. 25, the median value is $f_{d}\sim19$ for galaxies. Given that the cluster baryonic mass is dominated by the hot intra-cluster gas it is clear that this value of $f_{d} \sim 19$ for individual galaxies cannot account for the value of $f_{d} \sim 8$ for the cluster as a whole - confirming the long held notion that there must also be dark matter in the inter-galactic space.  There is $\sim 10^{13}$ M$_{\odot}$ Mpc$^{-3}$ of dark matter in the cluster as a whole, but only $\sim 10^{12}$ M$_{\odot}$ Mpc$^{-3}$ of this can be accounted for by the cluster galaxies.

\begin{figure}
\centering
\includegraphics[scale=0.52]{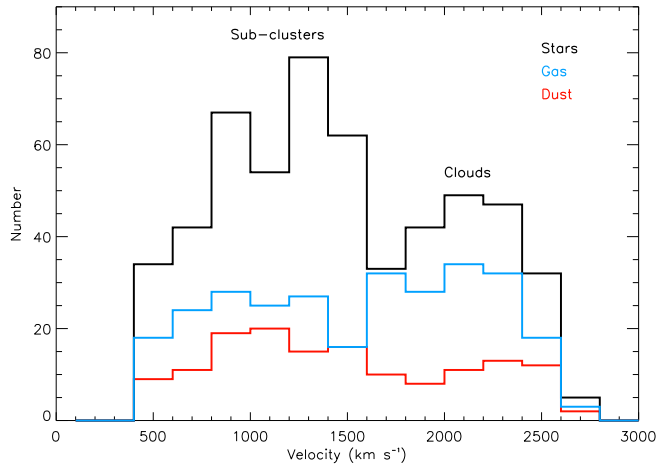}
\caption{The distribution of galaxy velocities for the three samples. Black 546 galaxies from VCC+, blue 261 galaxies from ALFALFA and red 146 galaxies selected from the Herschel sample. The bi-modal distribution is due to the cluster sub-structure, which can be split into the two main sub-clusters and two, probably infalling, clouds.} 
\end{figure} 

\begin{table*}
\begin{center}
\begin{tabular}{l|cccccc}
Sample & Number in sample & $<v>$ & $\sigma_{v}$ & $R_{e}$ & $M_{Cluster}$ & $\rho_{Dyn}$  \\
       &                  & (km s$^{-1}$) & (km s$^{-1}$) & (Mpc) & ($10^{14}$ $M_{\odot}$) & ($10^{12}$ $M_{\odot}$ Mpc$^{-3}$) \\ \hline
 Stars (VCC+)  & 546 & 1465 & 590 & 1.9 & 4.7 & 7.5 \\
 Gas (ALFALFA)  & 261 & 1552 & 616 & 1.8 & 4.7 & 7.5 \\
 Dust (Herschel) & 146 & 1475 & 616 & 2.2 & 5.7 & 9.2  \\
X-ray (Bohringer et al. 1994) & - & - & - & - & 4.2 & 6.7 \\
\end{tabular}
\caption{The dynamical properties of the cluster as measured by the various samples and compared with the total mass derived from X-ray observations.}
\end{center}
\end{table*}

\begin{figure}
\centering
\includegraphics[scale=0.52]{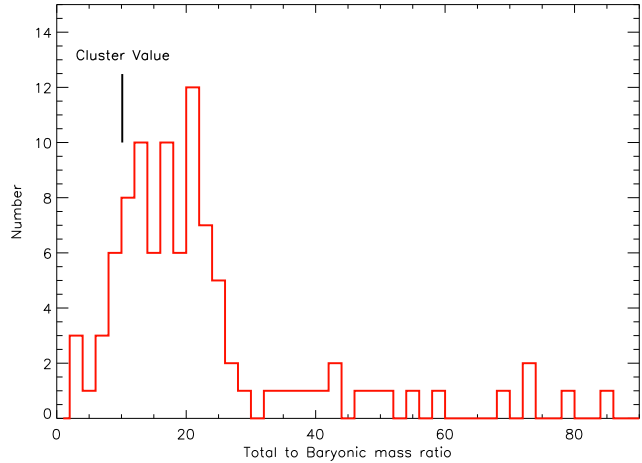}
\caption{The distribution of dynamical to baryonic mass for 100 Virgo cluster galaxies that have both a HI velocity width and a measured inclination. The black line marks the overall value for the cluster. } 
\end{figure}

\section{Some comments on the properties of the sub-clusters}
Finally, we will return to something we discussed in the introduction - the structure of the cluster. In this section we briefly compare the properties of the two sub-clusters, A and B, with the properties of the whole cluster that we have discussed earlier. The data is given in Table 8 and we will comment on each entry in this table. Each parameter has been calculated as described earlier in the text.

Binggeli et al. (1987) using optical data describe sub-cluster A as being rich in early type galaxies compared to sub-cluster B. In our Herschel sample we find no difference in the morphological mix of the two sub-clusters (Table 8). This is perhaps not surprising because early type galaxies will always be underrepresented in a sample selected by emission from the inter-stellar medium. 

We have obtained the masses of stars, gas and metals given in Table 8 by just summing the masses of all the 207 galaxies in the Herschel sample. This gives values for the cluster as a whole that are different by factors of 1.02, 1.59, 1.7 for stars, gas and dust respectively from those given in Table 5, which were obtained by integrating the mass functions (then using the density given in Table 5 multiplied by the volume of 62.4 Mpc$^{-3}$ to give the mass). We have adjusted the values given in Table 8 using the above factors so that they are consistent with Table 5. It is clear that sub-cluster A contains by far the majority of the mass, for example 85\% of the stellar mass is in sub-cluster A.

The ratios of mass of stars (M$_{Stars}$) to mass of gas (M$_{Gas}$) and mass of gas (M$_{Gas}$) to mass of metals (M$_{Metals}$) are both pretty consistent for both sub-clusters, though sub-cluster B is more gas rich compared to sub-cluster A. The mean effective yield ($p_{eff}$), as described in section 8.1, is the same for each sub-cluster indicating that their chemical evolutionary history is very much the same.

Both sub-clusters have about the same calculated mean velocity ($<v>$), but note the discussion in section 10 above about not including galaxies with velocities less than 400 km s$^{-1}$.  The similarity of mean velocities makes it difficult to distinguish sub-cluster membership without independent distance information (Gavazzi et al. 1999). Velocity dispersions ($\sigma_{v}$) are very similar for both the cluster and both the sub-clusters. Sub-cluster B is characterised by a smaller effective radius ($R_{e}$). The volumes (V) have been calculated, assuming spherical distributions and sizes of 1.2$R_{e}$ - section 10.

Not only does sub-cluster A dominate the baryonic mass it also dominates the dynamically measured mass (M$_{Dynamical}$) with 91\% of the total mass. Given the factor of 10 difference in stellar mass between sub-cluster A and B it is surprising that sub-cluster B has a dynamical mass only a factor of $\sim$0.7 smaller than A. This is reflected in the dynamical to baryonic mass ratio (M$_{Dynamical}$/M$_{Baryon}$, where M$_{Baryon}$ is the total mass in stars, gas, metals and X-ray gas), which is a factor of more than 10 larger than that for the cluster as a whole and sub-cluster A. The sub-cluster B dynamical mass is also discrepant when compared with the gravitating mass measured by the x-rays, which at $2 \times 10^{13}$ M$_{\odot}$ is more than a factor of 10 less than that measured by the dynamics. Using the gravitating mass calculated from the x-rays leads to a value of M$_{Dynamical}$/M$_{Baryon}=5$. However, Bohringer et al. (1994) note that the X-ray emission of M49, at the centre of sub-cluster B, is much more centrally condensed than that of the spatial distribution of galaxies in the sub-cluster and that the relation between gravitational mass and gas mass may be more like that of an isolated galaxy (Nulsen et al. 1984). In which case we may have under estimated M$_{X-ray}$ by as much as a factor of 100 making the problem even worse. It is difficult to know what is going on here, whether there is a problem, and if there is whether it is with M$_{Dynamical}$ or M$_{X-ray}$.

Finally, the star formation rate density has been calculated using the sum of the star formation rates in each sub-cluster and the volume (V). These rate densities do not vary greatly between the sub-clusters. Note that the SFR densities for the sub-clusters are less than that of Virgo because their combined volume is greater than that of Virgo.

\begin{table*}
\begin{center}
\begin{tabular}{cccc}
 & Virgo & Sub-cluster A & Sub-cluster B   \\ \hline
Number & 207 & 147 & 60 \\
Morphology & & & \\
\% type 1 & 12 & 12 & 12 \\
\% type 2 & 20 & 20 & 22 \\
\% type 3 & 27 & 26 & 28 \\
\% type 4 & 41 & 42 & 38 \\
M$_{Stars}$ (M$_{\odot}$) & $2.1 \times 10^{12}$ & $1.8 \times 10^{12}$ & $2.6 \times 10^{11}$ \\
M$_{Gas}$ (M$_{\odot}$) & $2.7 \times 10^{11}$ & $2.2 \times 10^{11}$ & $5.0 \times 10^{10}$ \\
M$_{Metals}$ (M$_{\odot}$) & $2.1 \times 10^{9}$ & $1.7 \times 10^{9}$ & $3.8 \times 10^{8}$ \\
M$_{X-rays}$ (M$_{\odot}$) & $8.4 \times 10^{13}$ & $7.6 \times 10^{13}$ & $4.0 \times 10^{12}$ \\
M$_{Stars}$/M$_{Gas}$ & 7.8 & 8.4 & 5.2 \\
M$_{Gas}$/M$_{Metals}$ & 129 & 128 & 132 \\
$p_{eff}$ & 0.009 & 0.009 & 0.009 \\
$<v>$ (km s$^{-1}$) & 1475 & 1515 & 1409 \\
$\sigma_{v}$ (km s$^{-1}$) & 616 & 609 & 629 \\
R$_{e}$ (Mpc) & 2.2 & 2.0 & 1.4 \\
V (Mpc$^{3}$) & 62.4 & 57.8 & 18.7 \\
M$_{Dynamical}$ (M$_{\odot}$) & $5.7 \times 10^{14}$ & $5.2 \times 10^{14}$ & $3.8 \times 10^{14}$ \\
M$_{Dynamical}$/M$_{Baryon}$  & 8 & 7 & 88 \\
SFR density (M$_{\odot}$ yr$^{-1}$ Mpc$^{-3}$) & 2.0 & 1.8 & 1.2 \\
\end{tabular}
\caption{The derived properties of the Virgo cluster compared to those of sub-clusters A and B as defined by galaxies detected by Herschel.}
\end{center}
\end{table*}

\section{Conclusions}
\begin{itemize}
\item Faint galaxy number counts both on and off the cluster field indicate that the optical selection of galaxies does not miss a population of previously undetected cluster far infrared sources.
\item Cluster luminosity functions are generally flatter at the faint end than those derived for field galaxies - the cluster lacks faint far-infrared sources.
\item The cluster dust mass function has a similar shape to that of field galaxies so the differences in the luminosity functions must be due to temperature - more low dust mass star forming galaxies in the field.
\item The cluster is over dense in dust by about a factor of 100 compared to the field (Table 2).
\item Individual galaxies have a range in global dust temperatures similar to that found in different regions of a typical galaxy like M31 (15-25 K).
\item Individual galaxies have the same $T-\beta$ relation as that found in different regions of a typical galaxy like M31.
\item We have used SDSS data to optically identify 43 new cluster members to add to those already listed in the VCC.
\item The cluster stellar mass function has a similar shape to that of the field and the cluster. The cluster is over dense in stars by about a factor of 100 compared to the field (Table 3).
\item We used ALFALFA data to identify 65 new HI detections that are not in the VCC.
\item The cluster atomic gas mass function is far less steep than that of the field. The cluster is over dense in atomic gas by only about a factor of 16 compared to the field - consistent with the loss of gas, but not stars and dust in the cluster environment (Table 4).
\item The mean metallicity of cluster galaxies is about 0.7$z_{\odot}$ and 50\% of the metals are in the gas phase with the rest residing in the dust.
\item We adjust the atomic gas masses for the contributions of He, H$_{2}$, and warm and hot gas to get the total gas mass. We use the observed values of (O/H) to measure the mass of metals in the gas phase.
\item The mass density of stars in galaxies is 8 times that of the gas and the gas mass density is 130 times that of the metals (Table 5).
\item We consider a chemical evolution model using the total mass of baryons in stars, gas and metals. We find no relation between the effective yield and mass, but as there is a mass metallicity relation we suggest that this relation may be more the consequence of age differences rather than a consequence of preferential mass loss by low mass galaxies.
\item The chemical evolution model predicts that the cluster galaxies have lost about 1.4 times their stellar mass to the intergalactic medium.
\item The effective yield depends on galaxy colour and morphological type - earlier type galaxies, not particularly ones of lower mass, seem to have lost more of their mass to the intergalactic medium.
\item Cluster galaxies appear to follow the same sSFR versus stellar mass relation as those in the field - lower mass galaxies have higher sSFR.
\item Lower mass late type galaxies have shorter star formation time scales and longer gas depletion time scales than more massive earlier types - they are middle aged compared to the old age of the early types.
\item The gas specific star formation rate is approximately constant for all galaxies.
\item Cluster galaxies follow the same baryonic Tully-Fisher relation as field galaxies.
\item Cluster galaxies follow the same mass size relation as field galaxies.
\item Given how well many galaxy properties scale with galaxy mass it is not easy to see how this can be accommodated within a hierarchical picture of galaxy formation. In addition it seems that the least massive galaxies are the youngest.
\item The presence of the intergalactic X-ray gas, at first sight just indicates that the galaxy formation process is very inefficient, however within any scenario it is difficult to account for the origin of the gas' metallicity and hence the origin of the gas itself. The X-ray gas is not consistent with gas loss from the cluster galaxies or with the production of metals within galactic stars.
\item The cluster is about 150 times more dense in baryons than the cosmologically measured value of $\Omega_{b}$.
\item The cluster is about 200 times more dense in all forms of matter than the cosmologically measured value of $\Omega_{m}$.
\item Dark matter in the individual galaxies cannot account for the total mass density of the cluster.
\item The two sub-clusters A and B, as detected by Herschel, are similar in their morphological mix and in most other respects. The main exception is mass, with the baryonic mass of sub-cluster A being about an order of magnitude larger than sub-cluster B. It is not clear whether either the dynamic or the X-ray data can provide a good estimate of the gravitational mass of sub-cluster B.
\end{itemize}

\vspace{0.5cm}
\noindent
{\bf ACKNOWLEDGEMENTS} \\

The Herschel spacecraft was designed, built, tested, and launched under a contract to ESA managed by the Herschel/Planck Project team by an industrial consortium under the overall responsibility of the prime contractor Thales Alenia Space (Cannes), and including Astrium (Friedrichshafen) responsible for the payload module and for system testing at spacecraft level, Thales Alenia Space (Turin) responsible for the service module, and Astrium (Toulouse) responsible for the telescope, with in excess of a hundred subcontractors.

PACS has been developed by a consortium of institutes led by MPE (Germany) and including UVIE (Austria); KU Leuven, CSL, IMEC (Belgium); CEA, LAM (France); MPIA (Germany); INAF-IFSI/OAA/OAP/OAT, LENS, SISSA (Italy); IAC (Spain). This development has been supported by the funding agencies BMVIT (Austria), ESA-PRODEX (Belgium), CEA/CNES (France), DLR (Germany), ASI/INAF (Italy), and CICYT/MCYT (Spain).

SPIRE has been developed by a consortium of institutes led by Cardiff University (UK) and including Univ. Lethbridge (Canada); NAOC (China); CEA, LAM (France); IFSI, Univ. Padua (Italy); IAC (Spain); Stockholm Observatory (Sweden); Imperial College London, RAL, UCL-MSSL, UKATC, Univ. Sussex (UK); and Caltech, JPL, NHSC, Univ. Colorado (USA). This development has been supported by national funding agencies: CSA (Canada); NAOC (China); CEA, CNES, CNRS (France); ASI (Italy); MCINN (Spain); SNSB (Sweden); STFC (UK); and NASA (USA).

This research has made use of the NASA/IPAC Extragalactic Database (NED) which is operated by the Jet Propulsion Laboratory, California Institute of Technology, under contract with the National Aeronautics and Space Administration. 

Funding for the SDSS and SDSS-II has been provided by the Alfred P. Sloan Foundation, the Participating Institutions, the National Science Foundation, the U.S. Department of Energy, the National Aeronautics and Space Administration, the Japanese Monbukagakusho, the Max Planck Society, and the Higher Education Funding Council for England. The SDSS Web Site is http://www.sdss.org/.
The SDSS is managed by the Astrophysical Research Consortium for the Participating Institutions. The Participating Institutions are the American Museum of Natural History, Astrophysical Institute Potsdam, University of Basel, University of Cambridge, Case Western Reserve University, University of Chicago, Drexel University, Fermilab, the Institute for Advanced Study, the Japan Participation Group, Johns Hopkins University, the Joint Institute for Nuclear Astrophysics, the Kavli Institute for Particle Astrophysics and Cosmology, the Korean Scientist Group, the Chinese Academy of Sciences (LAMOST), Los Alamos National Laboratory, the Max-Planck-Institute for Astronomy (MPIA), the Max-Planck-Institute for Astrophysics (MPA), New Mexico State University, Ohio State University, University of Pittsburgh, University of Portsmouth, Princeton University, the United States Naval Observatory, and the University of Washington.

This publication makes use of data products from the Two Micron All Sky Survey, which is a joint project of the University of Massachusetts and the Infrared Processing and Analysis Center/California Institute of Technology, funded by the National Aeronautics and Space Administration and the National Science Foundation.

We acknowledge the usage of the HyperLeda database (http://leda.univ-lyon1.fr). 

IDL is a postdoctoral researcher of the FWO-Vlaanderen (Belgium)

\vspace{0.5cm}
\noindent
\large
{\bf References} \\
\small
Aannestad P., 1975, ApJ, 200, 30 \\
Abazajian K. et al., ApJS, 182, 543  \\
Arimoto N. and Yoshi Y., 1987, A\&A, 173, 23 \\
Arnaboldi M., et al., 2002, AJ, 123, 760 \\
Arnaboldi M., et al., 2003, AJ, 125, 514 \\
Arnett D., 1991, In 'Frontiers of Stellar Evolution', Ed. D. Lambert, ASP conf. ser., 20, 389 \\
Asplund M. and Garcia Perez A., 2001, A\&A, 372, 601 \\
Auld R. et al., 2013, MNRAS, 428, 1880 \\
Avila-Reese V., Zavala J, Firmani C. and Hernandez-Toledo H., 2008, AJ, 136, 1340 \\
Baes M., et al., 2010, A\&A, 518, 53 \\
Bell E., McIntosh Daniel H., Katz N. and Weinberg M., 2003, ApJS, 149, 289 \\
Bendo et al., 2003, AJ, 125, 2361 \\
Bendo et al., 2012, MNRAS, 419, 1833 \\
Bianchi S., 2013, A\&A, 552, 89 \\
Bicay M. and Giovanelli R., 1987, ApJ, 321, 645  \\
Bigiel F., Leroy A., Walter F., Brinks E., de Blok W., Madore B. and Thornley M., 2008, AJ, 136, 2846 \\
Binggeli B., et al., 1985, AJ, 90, 1681  \\
Binggeli B. Tammann G. and Sandage A., 1987, AJ, 94, 251  \\
Binggeli B., Popescu C. and Tammann G., 1993, A\&ASS, 98, 275  \\
Binney J. and Tremaine S., 1994, In 'Galactic Dynamics', Pub. Princeton University Press, p. 610  \\
Bohringer et al., 1994, Nature, 368, 828  \\
Bohringer H., 2004, In 'Recycling Intergalactic and Interstellar Matter', IAU sym. Vol 217, p. 92, Ed. P. Duc, J. \\
Boselli et al., 2012, A\&A, 540, 54  \\
Boselli et al., 2011, A\&A, 528, 107  \\
Boselli et al., 2010, A\&A, 518, 61 \\
Boselli A. and Gavazzi G., 2006, PASP, 118, 517 \\
Briel U., Henry J. and Bohringer H., 1992, A\&A, 259, L31 \\
Calzetti, D. et al., 2010, ApJ, 714, 1256 \\
Cen R. and Ostriker J., 1999, ApJ, 514, 1 \\
Chung, A., van Gorkom J., Kenney J., Crowl H. and Vollmer B., AJ, 139, 1741 \\
Clemens M. et al., 2013, MNRAS, submitted \\
Clemens M., et al., 2010, A\&A, 518, 50  \\
Clements D., et al., 2010, A\&A, 518, L8 \\
Corbelli E., et al., 2012, A\&A, 542, 32 \\
Cortese L., et al., 2012, A\&A, 540, 52 \\
Cortese L., et al., 2010, A\&A, 518, 49 \\
Cortese L., Bendo G., Isaak K., Davies J. and Kent B., 2010, MNRAS, 403, 26 \\
Cortese L. et al., 2008, MNRAS, 383, 1519 \\
Cote, P., et al., 2004, AJSS, 153, 223 \\
Coupeaud A., et al., 2011, A\&A, 535, A124 \\
Dale D. et al., 2012, ApJ, 745, 95 \\
Davies J., et al., 2012, MNRAS, 419, 3505 \\
Davies J., 2012, arXiv:1204.4649 \\
Davies J., et al., 2011, MNRAS, 415, 1883  \\
Davies J., et al., 2010, A\&A, 518, 48 \\
Davies J., et al., 2004, MNRAS, 349, 922 \\
De Looze I., et al. 2010, A\&A, 2010, 518, 54 \\
Desert F. et al. 2008, A\&A, 481, 411 \\
di Serego Alighieri. et al., 2013, A\&A, 552, 8 \\
Disney M., Romano J., Garcia-Appado D., West A., Dalcanton J. and Cortese L., 2008, Nature, 455, 1082 \\
Draine B., et al., 2013, arXiv:1306.2304v1
Dressler A., 1980, ApJ, 236, 351 \\
Dunne L. et al., 2011, MNRAS, 417, 1510 \\
Dwek E., 1998, ApJ, 501, 643 \\
Eales S., et al., 2010, PASP, 122, 499 \\
Eales S., et al., 2013, in preparation \\
Edmunds M., 1990, MNRAS, 246, 678 \\
Feldmeier J., Giardullo R. and Jacoby G., 1998, ApJ, 503, 109 \\
Feldmeier J., et al., 2003, ApJS, 145, 65 \\
Ferguson H. and Sandage A., 1990, AJ, 100, 1 \\
Ferguson H., Tanvir N. and von Hippel T., 1998, Nature, 391, 461 \\
Ferrarese L. et al., 2012, ApJS, 200, 4 \\
Galametz et al., 2012, MNRAS, 425, 763 \\
Gavazzi G., Boselli A., Scodeggio M., Pierini D. and Belsole E., 1999, MNRAS, 304, 595 \\
Gavazzi, G. Boselli, A. Donati, A. Franzetti, P. and Scodeggio, M., 2003, A\&A, 400, 451  \\
Giovanelli, R. and Haynes M., 1985, ApJ, 292, 404 \\
Giovanelli, R., et al., 2005, AJ, 130, 2598 \\
Giovanelli, R., et al., 2007, AJ, 133, 2569 \\
Gordon et al., 2010, A\&A, 518, 89 \\
Griffin M., et al., 2010, A\&A, 518, 3 \\
Grossi M., et al., 2010, A\&A, 518, 52  \\
Gupta A., Mathur S., Krongold Y., Velton A., Veeraraghavan A. and Rasker R., 2012, ApJ, 756, 8 \\
Gurovich S., Freeman K., Jerjen H., Stavely-Smith L. and Puerari I., 2010, 140, 676 \\
Haffner L., et al., 2009, Rev.Mod.Phys., 81, 969 \\
Haynes M., Giovanelli R. Guido C., 1984, ARA\&A, 22, 445 \\
Howk C. and Consiglio M., 2012, ApJ, 759, 97 \\
Hughes T., et al., 2013, A\&A, 550, 115 \\
Ibar et al., 2010, MNRAS, 409, 38  \\
Iglesias-Paramo J., et al., 2006, ApJ, 164, 381 \\
Impey C., Bothun G. and Malin D., 1988, ApJ, 330, 891 \\
Kewley L. and Ellison S., 2008, ApJ, 681, 1183  \\
Kennicutt R., 1998, ApJ, 498, 541 \\
Kennicutt, R. C., Jr., et al., 2003, PASP, 115, 928  \\
Kent B., et al., 2007, ApJ, 665, 15  \\
Kent B., et al., 2009, ApJ, 691, 1595  \\
Komatsu  E., et al., 2011, ApJSS, 192, 18 \\
Kulkarni G., Rollinde E., Hennawi J. and Vangioni E., 2013, arXiv1301.4201 \\
Lagache G., Abergel A., Boulanger F. and Puget J., 1998, A\&A, 333, 709 \\
Lee H., McCall M. and Richer M., 2003, AJ, 2975, 2997 \\
Lewis I., et al., 2002, MNRAS, 334, 673 \\
Li A., (2004), In 'Penetrating bars through masks of cosmic dust : the Hubble tuning fork strikes a new note', Proc. of a conference held at Pilanesburg National Park (SA). Ed. by D. L. Block, I. Puerari, K. C. Freeman, R. Groess, and E. K. Block, ASSL, vol. 319. Dordrecht: Kluwer Academic Publishers, p.535 \\
Lis D. et al. 1998, ApJ, 509, 299 \\
Magnelli, B., et al., 2013, A\&A, 553, 132 \\
Magrini, L., et al., 2012, MNRAS, 427, 1075 \\
Matteucci F. and Gibson B., 1995, A\&A, 304, 11 \\
McCarthy I., Bower R. and Balogh M., 2007, MNRAS, 377, 1457 \\
McGaugh S., Schombert J., Bothun G. and de Blok W., 2000, ApJ, 533, 99 \\
McLaughlin D., 1999, ApJ, 512, L9 \\
Mei S., et al., 2007, ApJ, 655, 144  \\
Meny C., et al., 2007, A\&A, 468, 171 \\
Mestel L., 1963, MNRAS, 126, 553 \\
Meyer D., Jura M. and Cardelli J., 1998, ApJ, 493, 222 \\
Mei  S., et al., 2010, Bulletin of the American Astronomical Society, Vol. 42, p.514 \\
Mihos C., Harding P., Feldmeier J. and Morrison H., 2005, ApJ, 631, 41 \\
Mihos C., Janowiecki S., Feldmeier J., Harding J. and Morrison H., 2009, ApJ, 698, 1879 \\
Negrello M., et al., 2013, MNRAS, in press \\
Neugebauer G. et al., 1984, ApJ, 278, L1  \\
Nulsen P., Stewart G. and Fabian A., 1984, MNRAS, 208, 185 \\
Panter B., Jimenez R., Heavens A. and Charlot S., (2007), MNRAS, 378, 1550 \\
Pappalardo C et al., 2012, A\&A, 545, 75 \\
Peng Y., et al., 2010, ApJ, 721, 193 \\
Phillipps S., 2005, In 'The Structure and Evolution of Galaxies', Pub. John Wiley, p. 220 \\ 
Phillipps S., Driver S., Couch W. and Smith R., 1998, ApJ, 498, 119 \\
Pilbratt et al., 2010, A\&A, 518, 1 \\
Planck collaboration, 2011, A\&A, 536, A24 \\
Planelles S. et al., 2013, MNRAS, 431, 1487 \\
Poglitsch et al., 2010, A\&A, 518, 2 \\
Remy-Ruyer A. et al., 2013, A\&A, in press \\
Roberts S., et al., 2004, MNRAS, 352, 478 \\
Robotham A. and Driver S., 2011, MNRAS, 413, 257 \\
Sabatini et al.,  2003, MNRAS, 341, 981  \\
Salpeter E., 1955, ApJ, 121, 161 \\
Sanders D., Mazzarella J., Kim D., Surace J. and Soifer B., 2003, AJ, 126, 1607  \\
Schmidt M., 1959, ApJ, 129, 243 \\
Schiminovich D. et al., 2007, ApJS, 173, 315 \\
Seki J. and Yamamota T., 1980, Ap\&SS, 72, 79 \\
Shapley H. and Ames A., 1926, Harvard Circ. No 294 \\
Shetty R., Kauffmann J., Schnee S. and Goodman A., 2009a, ApJ, 696, 676 \\
Shetty R., Kauffmann J., Schnee S., Goodman A. and Ercolano B., 2009b, ApJ, 696, 676 \\
Skibba R., et al., 2011, ApJ, 738, 123 \\
Skrutskie M., et al., 2006, AJ, 131, 1163 \\
Smith H., 1981, AJ, 86, 998 \\
Smith M., et al., 2010, A\&A, 518, 51 \\
Smith M., et al., 2012, ApJ, 756, 40 \\
Smith M., et al., 2012a, ApJ, 748, 123 \\\
Stark D., McGaugh S. and Swaters R., 2009, AJ, 138, 392 \\
Stepnik B. et al., 2003, A\&A, 398, 551 \\
Strong A. and Mattox J., 1996, A\&A, 308, 421 \\
Taylor R., 2010, PhD thesis, Cardiff University, UK. \\
Tremonti C., et al., 2004, ApJ, 613, 913 \\
Trenti M. and Stiavelli M., 2009, ApJ, 694, 879 \\
Tripp T., Savage B. and Jenkins E., ApJ, 534, 1 \\
Urban O., Werner N., Simionescu A., Allen S. and Bohringer H., 2011, MNRAS, 414, 210 \\
Vikhlinin A., et al., 2006, ApJ, 640, 691 \\
Vila-Costas M. and Edmunds M., 1992, MNRAS, 259, 121  \\
Warren S. et al., 2007, MNRAS, 375, 213 \\
White S., Navarro J., Evard A. and Frenk C., 1993, Nat, 366, 429 \\
Whittet D., 1991, 'Dust in the galactic Environment', IOP publishing, Bristol. \\
Young L., et al., 2011, MNRAS, 414, 940 \\

\pagebreak

\begin{table}
\begin{center}
{\bf Appendix 1}
\end{center}
\begin{center}
\begin{tabular}{lcccc}
Name   & RA & Dec & $v_{Helio}$           & $g$    \\ 
       & (2000)   &   (2000) & (km s$^{-1}$) & (mag)  \\ \hline
VCCA1  &  12 11  40.3 & 12 58  24.6 & 2220 & 18.3     \\
VCCA2  &  12 11  45.9 & 13 17  07.9  & 2466 & 18.2   \\
VCCA3  &  12 11  53.9 & 13 48  30.3 & 2016 & 17.3   \\
VCCA4  &  12 11  59.5 & 05 55  02.7   & 759  & 17.2     \\
VCCA5  &  12 12  40.7 & 06 50  27.8  & 2286 & 17.7   \\
VCCA6  &  12 14  23.5 & 06 45  51.0  & 2154 & 16.2     \\
VCCA7  &  12 14  28.2 & 05 54  31.3  & 1944 & 16.8     \\
VCCA8  &  12 14  44.6 & 12 47  22.7 & 2292 & 17.9   \\
VCCA9  &  12 14  44.9 & 06 09  16.2  & 2064 & 17.6   \\
VCCA10 &  12 15  17.1 & 06 23  35.4  & 2136 & 17.6   \\
VCCA11 &  12 16  10.1 & 15 07  25.3 & 570  & 17.5     \\
VCCA12 &  12 16  20.6 & 14 46  26.1 & 525  & 17.9  \\
VCCA13 &  12 17  44.8 & 06 04  57.6  & 2478 & 17.1     \\
VCCA14 &  12 17  46.6 & 05 22  36.2  & 2013 & 18.0     \\
VCCA15 &  12 18  04.9 & 07 44  45.0  & 2166 & 18.3  \\
VCCA16 &  12 18  20.6 & 04 51  13.6  & 2100 & 16.7  \\
VCCA17 &  12 18  28.3 & 06 04  33.9  & 2250 & 18.3    \\
VCCA18 &  12 19  27.8 & 04 34  42.6  & 1584 & 16.1    \\
VCCA19 &  12 19  28.5 & 06 16  22.6  & 1530 & 17.8    \\
VCCA20 &  12 19  49.2 & 05 11  06.1   & 1968 & 16.2  \\
VCCA21 &  12 19  53.7 & 06 01  56.4  & 1992 & 18.1  \\
VCCA22 &  12 20  30.9 & 06 38  23.8  & 1497 & 18.2  \\
VCCA23 &  12 20  35.0 & 05 48  27.3  & 2190 & 17.1  \\
VCCA24 &  12 20  56.3 & 05 11  20.5  & 2355 & 18.1  \\
VCCA25 &  12 21  22.5 & 05 56  51.7  & 1785 & 17.0  \\
VCCA26 &  12 22  04.7 & 07 44  22.0  & 1344 & 17.2    \\  
VCCA27 &  12 22  43.6 & 05 27  23.0  & 2310 & 18.2  \\
VCCA28 &  12 22  59.6 & 08 01  51.4  & 648  & 17.0    \\ 
VCCA29 &  12 23  03.2 & 05 47  09.5   & 2064 & 18.2    \\ 
VCCA30 &  12 25  04.2 & 15 42  40.7 & 1398 & 17.0  \\
VCCA31 &  12 25  31.5 & 11 09  30.1 & 906  & 17.3    \\ 
VCCA32 &  12 25  51.2 & 07 47  14.9  & 660  & 18.3  \\
VCCA33 &  12 26  24.9 & 10 34  54.8 & 1026 & 18.7  \\
VCCA34 &  12 26  47.9 & 07 40  17.6  & 618  & 15.7  \\
VCCA35 &  12 26  49.1 & 09 34  27.9  & 867  & 14.1    \\ 
VCCA36 &  12 29  14.6 & 07 52  39.1  & 1530 & 15.8  \\
VCCA37 &  12 29  51.2 & 14 03  59.3 & 1518 & 19.8  \\
VCCA38 &  12 30  02.6 & 09 24  11.8  & 924  & 17.4    \\
VCCA39 &  12 31  52.9 & 12 15  59.1 & 969  & 18.0  \\
VCCA40 &  12 33  40.3 & 12 44  13.6 & 1146 & 18.5  \\
VCCA41 &  12 33  44.7 & 10 59  39.8 & 1161 & 17.4  \\
VCCA42 &  12 33  56.4 & 07 42  25.8  & 828  & 18.0    \\  
VCCA43 &  12 37  34.3 & 08 28  55.2  & 1527 & 18.3  \\
\end{tabular}
\end{center}

\begin{center}
{\bf Table A1.} Virgo Cluster Catalogue Additional (VCCA) galaxies detected using SDSS data.
\end{center}
\end{table}

\begin{table}
\begin{center}
{\bf Appendix 2}
\end{center}
\begin{center}
\begin{tabular}{lccccc}

Name   & RA & Dec & $v_{Helio}$  &  $W_{50}$  & Line Flux   \\ 
       & (2000) & (2000) & (km s$^{-1}$) & (km s$^{-1}$) & (Jy km s$^{-1}$) \\ \hline
VCCA44   & 12 10 38.2 & 13 01 22.0 & 2394 &  34 &  0.56 \\
VCCA45   & 12 11 59.8 & 05 54 53.0 &  752 &  33 &  0.39 \\
VCCA46   & 12 12 01.6 & 10 24 08.0 & 1637 &  69 &  0.87 \\
VCCA47   & 12 12 59.2 & 07 18 03.0 & 2162 &  71 &  0.67 \\
VCCA48   & 12 13 10.1 & 13 34 32.0 & 2100 &  94 &  0.69 \\
VCCA49   & 12 13 41.8 & 12 53 52.0 & 2235 &  53 &  1.30 \\
VCCA50   & 12 13 49.8 & 05 21 01.0 & 1690 &  49 &  0.99 \\
VCCA51   & 12 14 11.7 & 12 47 21.0 &  613 &  90 &  0.68 \\
VCCA52   & 12 14 13.7 & 08 54 22.0 & 1933 &  95 &  2.07 \\
VCCA53   & 12 14 26.7 & 05 55 01.0 & 1921 & 110 &  0.96 \\
VCCA54   & 12 14 41.4 & 12 46 43.0 & 2279 &  44 &  0.39 \\
VCCA55   & 12 14 55.8 & 09 40 03.0 & 1693 &  31 &  0.34 \\
VCCA56   & 12 15 28.2 & 10 31 14.0 & 1990 &  59 &  0.83 \\
VCCA57   & 12 16 12.8 & 08 21 59.0 &  862 &  21 &  0.35 \\
VCCA58   & 12 16 27.5 & 06 03 00.0 & 2029 &  79 &  1.94 \\
VCCA59   & 12 16 34.3 & 10 12 35.0 & 2072 &  38 &  0.91 \\
VCCA60   & 12 17 27.1 & 12 55 49.0 & 2056 &  66 &  0.62 \\
VCCA61   & 12 17 27.4 & 07 19 46.0 & 2202 & 137 &  1.68 \\
VCCA62   & 12 17 33.8 & 14 23 47.0 & 2111 &  64 &  0.60 \\
VCCA63   & 12 17 49.1 & 15 04 52.0 & 2200 &  40 &  0.54 \\
VCCA64   & 12 17 55.5 & 14 44 45.0 & 1990 & 128 &  2.30 \\
VCCA65   & 12 17 59.7 & 08 09 50.0 & 2000 &  98 &  0.87 \\
VCCA66   & 12 18 05.1 & 14 45 16.0 & 1763 &  47 &  0.48 \\
VCCA67   & 12 18 11.1 & 04 38 13.0 & 2073 &  37 &  0.39 \\
VCCA68   & 12 18 47.7 & 04 48 12.0 & 1822 & 103 &  0.80 \\
VCCA69   & 12 19 16.8 & 06 15 21.0 & 1983 &  51 &  0.84 \\
VCCA70   & 12 19 18.8 & 06 24 10.0 & 1969 &  44 &  1.60 \\
VCCA71   & 12 19 20.4 & 12 57 10.0 & 2164 &  57 &  0.77 \\
VCCA72   & 12 19 30.5 & 06 00 45.0 & 1622 &  52 &  0.60 \\
VCCA73   & 12 19 42.5 & 05 35 15.0 & 2381 &  56 &  0.58 \\
VCCA74   & 12 19 57.8 & 05 26 01.0 & 2474 &  62 &  0.88 \\
VCCA75   & 12 20 37.5 & 14 36 08.0 &  587 &  29 &  0.31 \\
VCCA76   & 12 20 40.5 & 05 06 09.0 & 1691 &  74 &  0.71 \\
VCCA77   & 12 20 41.5 & 05 54 30.0 &  978 &  48 &  0.48 \\
VCCA78   & 12 20 49.7 & 05 58 60.0 & 1772 &  46 &  0.32 \\
VCCA79   & 12 21 12.7 & 10 37 19.0 & 2606 &  37 &  0.51 \\
VCCA80   & 12 22 33.7 & 04 40 28.0 & 2214 & 103 &  0.89 \\
VCCA81   & 12 23 17.8 & 05 36 19.0 & 1787 & 178 &  0.90 \\
VCCA82   & 12 23 57.5 & 07 26 57.0 & 1229 &  55 &  0.66 \\
VCCA83   & 12 24 04.7 & 08 17 35.0 & 1370 &  56 &  1.13 \\
VCCA84   & 12 24 48.6 & 07 54 23.0 &  797 &  20 &  0.31 \\
VCCA85   & 12 24 51.0 & 04 03 17.0 & 1771 &  34 &  0.37 \\
VCCA86   & 12 25 32.4 & 11 09 13.0 &  900 &  48 &  0.36 \\
VCCA87   & 12 26 01.8 & 08 10 18.0 & 1304 &  33 &  1.14 \\
VCCA88   & 12 26 19.4 & 12 53 30.0 & 2246 & 135 &  2.25 \\
VCCA89   & 12 27 13.7 & 07 38 18.0 & 1179 &  44 &  1.28 \\
VCCA90   & 12 29 30.4 & 08 47 12.0 &  539 &  36 &  0.35 \\
VCCA91   & 12 29 42.8 & 09 41 54.0 &  524 & 116 &  1.24 \\
VCCA92   & 12 29 58.3 & 08 26 04.0 &  609 &  53 &  0.49 \\
VCCA93   & 12 30 19.4 & 09 35 18.0 &  603 & 252 &  2.84 \\
VCCA94   & 12 30 25.8 & 09 28 01.0 &  488 &  62 &  2.74 \\
VCCA95   & 12 30 44.4 & 05 50 06.0 & 2346 &  96 &  0.82 \\
VCCA96   & 12 30 44.4 & 05 52 11.0 & 2332 &  52 &  0.57 \\
VCCA97   & 12 31 19.0 & 09 27 49.0 &  607 &  56 &  0.76 \\
VCCA98   & 12 31 26.7 & 09 18 52.0 &  480 &  53 &  0.97 \\
VCCA99   & 12 32 08.2 & 05 50 11.0 & 1838 &  39 &  0.59 \\
VCCA100  & 12 32 23.8 & 05 54 37.0 & 1830 &  50 &  0.66 \\
VCCA101  & 12 32 36.5 & 06 00 58.0 & 1799 & 112 &  0.90 \\
VCCA102  & 12 33 08.4 & 05 51 47.0 & 1917 &  32 &  0.92 \\
VCCA103  & 12 33 15.5 & 05 02 19.0 & 1825 &  29 &  0.76 \\
VCCA104  & 12 33 34.5 & 06 02 30.0 & 1872 & 104 &  1.05 \\
VCCA105  & 12 33 36.9 & 06 26 43.0 & 1804 & 103 &  2.19 \\
VCCA106  & 12 34 02.6 & 05 57 49.0 &  648 &  21 &  0.43 \\
VCCA107  & 12 34 45.3 & 12 46 49.0 &  785 & 145 &  0.95 \\
VCCA108  & 12 35 06.9 & 02 30 51.0 & 1692 & 101 &  1.82 \\
VCCA109  & 12 35 23.1 & 05 02 32.0 & 1805 &  37 &  0.46 \\

\end{tabular}
\end{center}

\begin{center}
{\bf Table A2.} Virgo Cluster Catalogue Additional (VCCA) galaxies detected at 21cm using ALFALFA data.
\end{center}
\end{table}

\end{document}